\documentclass[%
 reprint,
superscriptaddress,
 amsmath,amssymb,
 aps,
pra,
]{revtex4-1}
\pdfoutput=1

\usepackage{amsmath}
\usepackage{amssymb}
\usepackage{graphicx}% Include figure files
\usepackage{epstopdf}

\usepackage{url}
\usepackage{hyperref}
\usepackage{breakurl}

\usepackage{color}
\usepackage[usenames,dvipsnames]{xcolor}
\hypersetup{breaklinks=true, colorlinks=true, linkcolor=BrickRed, urlcolor=blue!50!black, citecolor=blue!50!black}
\DeclareMathOperator{\sech}{sech}

\begin{document}
\preprint{APS/123-QED}
\title{Driven Lorentz model in discrete time}

\author{Dan Shafir}
\email{dansh5d@gmail.com}
\affiliation{Physics Department, Bar-Ilan University, Ramat Gan 5290002, Israel}
\author{Alessio Squarcini}
\email{alessio.squarcini@uibk.ac.at}
\affiliation{Institut f\"ur Theoretische Physik, Universit\"at Innsbruck, Technikerstra{\ss}e 21A, A-6020 Innsbruck, Austria}
\author{Stanislav Burov}
\email{stasbur@gmail.com}
\affiliation{Physics Department, Bar-Ilan University, Ramat Gan 5290002, Israel}
\author{Thomas Franosch} 
\email{thomas.franosch@uibk.ac.at}
\affiliation{Institut f\"ur Theoretische Physik, Universit\"at Innsbruck, Technikerstra{\ss}e 21A, A-6020 Innsbruck, Austria}

%\pacs{PACS}

\begin{abstract}
We consider a tracer particle performing a random walk on a two-dimensional lattice in the presence of immobile hard obstacles. Starting from equilibrium, a constant force pulling on the particle is switched on, driving the system to a new stationary state. Our study calculates displacement moments in discrete time (number of steps $N$) for an arbitrarily strong constant driving force, exact to first order in obstacle density. 
We find that for fixed driving force $F$, the approach to the terminal discrete velocity scales as $\sim N^{-1} \exp(- N F^2 / 16)$ for small $F$, differing significantly from the $\sim N^{-1}$ prediction of linear response. 
Besides a non-analytic dependence on the force and breakdown of Einstein's linear response, our results show that
fluctuations in the directions of the force are enhanced in the presence of obstacles. Notably, the variance grows as 
$\sim N^3$ (superdiffusion) for $F \to \infty$ at intermediate steps, reverting to normal diffusion ($\sim N$) at larger steps, a behavior previously observed in continuous time but demonstrated here in discrete steps for the first time. Unlike the exponential waiting time case, the superdiffusion regime starts immediately at $N=1$.
The framework presented allows considering any type of waiting-time distribution between steps and transition to continuous time using subordination methods. 
Our findings are also validated through computer simulations.
\end{abstract}

\maketitle

\newcommand{\avg}{\langle \tau_i \rangle}
\newcommand{\fr}{\frac}
\newcommand{\tl}{\tilde}

\section{Introduction}

The transport of molecules, colloids and particulate matter in disordered media is ubiquitous in natural, industrial and technological processes. Such systems have been extensively studied using a random-walk-on-a-lattice approach. One of the best-known examples is the continuous-time random walk~\cite{scher1975anomalous, shlesinger1974asymptotic, klafter1980derivation, scher1991time} (CTRW), popularized by Montrol and Scher~ \cite{scher1975anomalous} to model charge transport in amorphous materials and later used very successfully for the description of transport in porous and biological mediums~\cite{metzler2022modelling,waigh2023heterogeneous,nissan2018inertial,metzler2014anomalous,hofling2013anomalous,weigel2011ergodic}. The idea behind CTRW is to build upon the classical random walk, which is a succession of random steps, by introducing a waiting time distribution between steps, i.e., the medium is a form of energetic landscape which gives rise to waiting or trapping times. It has been shown that when these waiting times follow a scale-free distribution where the mean waiting time diverges, it leads to intriguing phenomena such as anomalous diffusion, weak ergodicity breaking and aging, seen in single quantum dots~\cite{stefani2009beyond}, transport of biomolecules inside the cell~\cite{Barkai2012single, jeon2011vivo, tabei2013intracellular, hofling2013anomalous, he2008random, sokolov2008statistics, metzler2014anomalous} and glassy systems~\cite{ shafir2022case, bouchaud1992weak, monthus1996models, rinn2000multiple, rinn2001hopping, berthier2011theoretical} just to name a few.

A second prominent model is the lattice Lorentz gas~\cite{lorentz1905mouvement}, describing obstructed transport in heterogeneous environments such as the crowded world inside biological cells. The model consists of a tracer particle performing random walk on a lattice where a fraction of the sites (density) is occupied by immobile hard obstacles. These obstacles, placed at random lattice locations, are treated with reflecting boundary conditions. 
Many works probe the characteristics of the Lorentz model system by studying the response of the particle to an external driving force~\cite{leitmann2018time, leitmann2017time, benichou2016nonlinear, benichou2014microscopic, illien2014velocity, jack2008negative, Squarcini_2024_1, Squarcini_2024_2}.
The emphasis is usually for a near-neighbor hopping process where the average waiting time between jumps is finite (often an exponential distribution). Already for finite average waiting time between hops, the presence of obstacles alters the dynamics of the driven system in a non-trivial manner since repeated collisions of the tracer with obstacles introduce correlations and persistent memory~\cite{jack2008negative, leitmann2013nonlinear, basu2014mobility, baiesi2015role, illien2014velocity, benichou2014microscopic, illien2015distribution, benichou2016nonlinear, hofling2013anomalous}. 
For example, in Ref.~\cite{leitmann2013nonlinear}, a first-order expansion in the density of immobile obstacles already presents a surprising force-dependent exponential decay towards the steady-state drift velocity; In contrast, a very dense system study reveals a surprising very short lived initial high velocity value that abruptly drops to a terminal 'low' value \cite{illien2014velocity}.

The combination of obstacles with different types of distributions of waiting times between steps -- which we refer to as temporal disorder, has received limited attention so far. One instance of this is a power law distribution resulting in diverging mean waiting times.
In our work, we find the moments of displacement in the domain of number of steps, i.e. discrete time, for arbitrarily strong constant driving force, accurate to first order in the obstacle density. This theoretical approach will establish a framework for future research that will allow to consider any type of temporal disorder and transition to continuous time by means of the method of subordination \cite{barkai2001fractional,meerschaert2004limit,sokolov2005diffusion,yuste2005trapping,saichev1997fractional}; i.e., a summation of conditional probability on all the possible outcomes of the number of steps during total time $t$ of the process. 
Our solution technique employs a scattering formalism borrowed from quantum mechanics~\cite{ballentine2014quantum} that has been successfully applied to the analytic study of the driven lattice Lorentz gas in continuous time with an exponential distribution between steps~\cite{leitmann2013nonlinear, leitmann2018time}. 

In this work, the discrete nature of working in number of steps leads to different mathematical challenges and results compared to the continuous case, since we will be dealing mainly with summation techniques (generating functions) instead of integrals (Laplace transforms). We show that for fixed driving $F$, the step-dependent approach ($N$ discrete time) towards the terminal discrete velocity (average position divided by steps) is exponentially fast rather than a power law decay as predicted by linear response.
Furthermore, we provide the first and second moment to first order in the obstacle density. Showing the dependence on the force is complex and non-analytic, indicating breaking of Einstein's linear response even for small forces.
The intuitive picture is that obstacles suppress the fluctuations in the direction of the force. Our results indicate that for forces large enough, increasing disorder leads to an enhancement. We show that there is a window at intermediate values of steps $N$ where the variance grows as $\sim N^{\alpha}$ with a true exponent of $\alpha=3$ in the limit $F \to \infty$ which then drops to normal diffusion ($\alpha=1$) at large steps. Such intriguing behavior has been already documented for the lattice Lorentz gas~\cite{illien2018nonequilibrium, leitmann2017time, illien2014velocity}. We provide the first evidence that such an anomalous behavior occurs also in the discrete domain of the number of steps. Our findings are validated by high-precision stochastic simulations.

As a consequence of the approach developed in this paper, our framework will find natural application to study the impact of heavy-tailed distribution of waiting times with a diverging average in the presence of obstacles.
Another example that deserves to be considered in the presence of obstacles is the case of a temporal quenched disorder which leads to correlations and memory effects~\cite{akimoto2018non,akimoto2020trace,shafir2022case, burov2011time,burov2017quenched}. A quenched temporal disorder by itself is known to exhibit surprising effects, including mobility enhancement in confined geometries~\cite{shafir2024disorder} and non self-averaging leading to universal fluctuations of diffusivity~\cite{ akimoto2016universal}.
Such interpretations which combine temporal and obstacle disorder may be a more complete picture to describe transport in a wide variety of systems and can lead to new theoretical advancements in the field.

\section{The model and solution technique} \label{sec:the_model}
\begin{figure}[t]
\centering
\includegraphics[width=0.45\textwidth]{"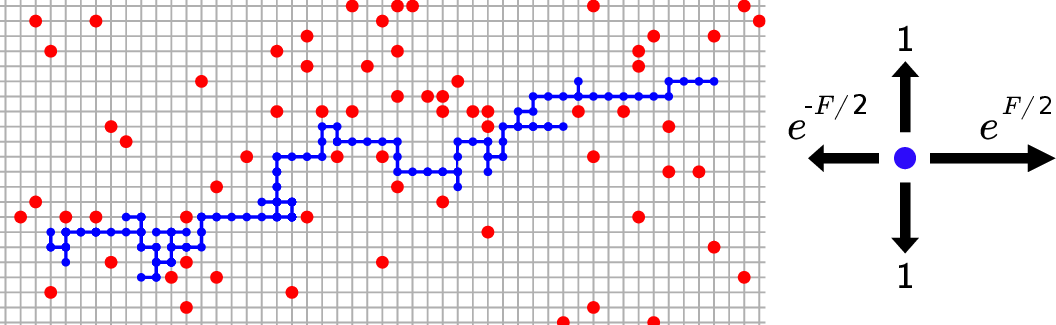"}
\caption{Left panel: A trajectory of a tracer pulled to the right by a constant force in the presence of hard obstacles (red sites) in the driven lattice Lorentz gas.
If the tracer attempts to jump to an obstacle site, it remains in the same position as just before the jump.
Right panel: relative transition probabilities with respect to a jump perpendicular to the force in the driven lattice Lorentz gas.
}
\label{Fig:1}
\end{figure}

In the two-dimensional lattice Lorentz gas, a tracer particle performs a random walk on a square lattice of size $L \times L$ (where $L \in \mathbb{N}$) defined by the collection of sites $\mathbf{r} \in \Lambda=\{(x, y) \in \mathbb{N} \times \mathbb{N} : (1 \leq x , y \leq L) \}$.
Here the lattice spacing $a$ is set to unity for convenience such that the lattice sites assume only integer values and the total number of steps performed is $N$.
We assume that the number of sites $L^2$ is very large and approaching the limit $L \to \infty$, i.e., the thermodynamic limit.
At the boundaries we employ periodic boundary condition.
In every step the tracer performs a nearest-neighbor jump of size $\mathbf{d} \in \mathcal{N}=\left\{\pm \mathbf{e}_x, \pm \mathbf{e}_y\right\}$ where $\mathbf{e}_x$ and $\mathbf{e}_y$ are perpendicular unit vectors in the $x$ and $y$ direction respectively. 
The lattice consists of free sites accessible to the tracer as well as sites with randomly placed immobile
hard obstacles of density $n$ (fraction of excluded sites). If the tracer attempts to jump onto an obstacles site, it remains at its initial position before the jump but still the counter for the discrete time is increased by one.
At zero steps (discrete time), a constant force acting on the tracer is switched on, and we use the thermal equilibrium state in the absence of driving $F = 0$ as the initial condition, i.e., the particle is equally likely to be anywhere on the $L^2$ sites at 'time' $N=0$.
The external force pulls the tracer only along the $x$ direction of the lattice and the strength of the force is characterized by the dimensionless force $F= \text{force}  \times a / k_B T$, where $k_B$ is Boltzmann's constant and $T$ is the temperature. 
The transition probabilities for a single step $W(\mathbf{d})$ obey detailed balance, $W\left(\mathbf{e}_x\right) / W\left(-\mathbf{e}_x\right)=e^F$ in the $x$ direction and $W\left(\mathbf{e}_y\right) / W\left(-\mathbf{e}_y\right)=1$ in the $y$ direction. 
Correspondingly, we choose the transition probabilities parallel and perpendicular to the applied force as
\begin{eqnarray}
W\left(\pm \mathbf{e}_x\right)=\Gamma e^{\pm F / 2},
\end{eqnarray}
and
\begin{eqnarray}
W\left(\pm \mathbf{e}_y\right)=\Gamma,
\end{eqnarray}
respectively as depicted in Fig.~\ref{Fig:1} where
\begin{eqnarray} \label{eq:transition_normalization}
    \Gamma = 1 / \left(e^{F / 2}+e^{-F / 2}+2\right),
\end{eqnarray}
is the normalization factor.
%%%%%%%%%%%%%%%%%%%%%%%%%%%%
%%%%%%%%%%%%%%%%%%%%%%%%%%%%%

We use an approach similar to Ref. \cite{leitmann2013nonlinear, leitmann2018time} to find the propagator (defined below) which will enable us to derive the moments of the displacement as a function of the number of steps $N$.
We first begin for the case of no obstacles, i.e., the free system, since it is more simple and to establish our formalism.
It is convenient to exploit the analogy of the master equation to a Schr\"odinger equation. 
Hence we consider the Hilbert space of lattice functions $\Lambda \rightarrow \mathbb{C}$ spanned by the orthonormal basis of position kets $|\mathbf{r}\rangle$. 
We denote by  $p_N(\mathbf{r})$  the probability     for the tracer to be at $\mathbf{r}$ after $N$ jumps and define an abstract ket $|p_N \rangle := \sum_{\mathbf{r}\in \Lambda} p_N(\mathbf{r}) |\mathbf{r}\rangle$. The probabilities can thus be obtained as $p_N(\mathbf{r}) = \langle \mathbf{r} | p_N \rangle$.
We assume that the force is switched on at step $N=0$ such that the thermal equilibrium state $|p_0 \rangle$ evolves towards a 
new stationary state $| p_{\text{st}} \rangle = \lim_{N\to\infty} | p_N \rangle$ indicated by the subscript "st".
We define the state of the system at step $N=0$ to be the equilibrium state of the empty lattice with no driving, $\langle \mathbf{r} | p_0 \rangle = 1/L^2$, meaning the probability to find the tracer at any of the $L^2$ sites of the lattice is uniform.
We define the single jump matrix for the free system with no obstacles indicated by the subscript $0$ as
%%%%%%%%%%%%%%%%%%%%%%%%%%%%%%5
\begin{equation} \label{eq:single_step_propagator}
\hat{M}_0=\sum_{\mathbf{r}} \sum_{\mathbf{d} \in \mathcal{N}} W(\mathbf{d})|\mathbf{r}\rangle\langle\mathbf{r}-\mathbf{d}|,
\end{equation}
%%%%%%%%%%%%%%%%%%%%%%%%%%%%%%
which entails the rule to propagate probabilities in time by $| p_{N+1} \rangle = \hat{M}_0 | p_{N} \rangle$.
Here the matrix element $\langle \mathbf{r} | \hat{M}_0 | \mathbf{r'} \rangle$ is the transition probability from $\mathbf{r'}$ to $\mathbf{r}$.
Now we define the free propagator to be the $z$-transform (sometimes called generating function) \cite{weiss1994aspects, hughes1995random, klafter2011first} of the single jump matrix $\hat{M}_0$ 
%%%%%%%%%%%%%%%%%%%%%%%%%%
\begin{equation}
    \hat{G}_0(z) = \sum_{N=0}^{\infty} (\hat{M}_0)^N z^N.
\end{equation}
%%%%%%%%%%%%%%%%%%%%%%%%%%%
It is convenient to perform a spatial Fourier transform, defined by
%%%%%%%%%%%%%%%%%%%%%%%%%%%%%%%%%%%%%%%%%
\begin{equation} \label{eq:base_k_definition}
    |\mathbf{k}\rangle=\frac{1}{L} \sum_{\mathbf{r} \in \Lambda} \exp (\mathrm{i} \mathbf{k} \cdot \mathbf{r})|\mathbf{r}\rangle ,
\end{equation}
%%%%%%%%%%%%%%%%%%%%%%%%%%%%%%%%%%%%%%%%
where $\mathbf{k}=\left(k_x, k_y\right) \in \Lambda^*=\{(2 \pi n_x / L, 2 \pi n_y / L): (n_x, n_y) \in \Lambda\}$ and $\mathbf{k} \cdot \mathbf{r}=k_x x+k_y y$. 
Since the matrix $\hat{M}_0$ is translationally invariant, in the plane wave basis becomes diagonal,  i.e., $\langle \mathbf{k} | \hat{M}_0 | \mathbf{k}' \rangle = \lambda (\mathbf{k}) \delta_{\mathbf{k}, \mathbf{k}'}$ with $\lambda(\mathbf{k})$ the eigenvalues, sometimes referred to as the characteristic function~\cite{klafter2011first, hughes1995random, weiss1994aspects}.
The jumps are independent of each other thus the free propagator $\hat{G}_0$ is also diagonal in $k$-space,
\begin{eqnarray} \label{eq:G0_k-space}
G_0(z, \mathbf{k}) &=& \langle \mathbf{k} | \hat{G}_0(z) | \mathbf{k} \rangle\\ \nonumber
&=& \sum_N [\lambda(\mathbf{k})]^N z^N = [1-z \lambda(\mathbf{k})]^{-1}.
\end{eqnarray}
The function $G_0(z, \mathbf{k})$ is also called the moment generating function since $(-i)^m\partial^m G_0(z, \mathbf{k}) / \partial k_x^m  |_{\mathbf{k=0}}$ is  the $m$-th moment of displacement along the $x$ axis (parallel to the direction of the applied force) \cite{klafter2011first, hughes1995random, weiss1994aspects}.
This will provide the solution for the first and second moments in the obstacle-free system, a similar approach will be taken if the general propagator for the case of obstacles $G(z,\mathbf{k})$ is known.
Thus the solution strategy of finding the moments will be achieved by finding the general propagator $G(z,\mathbf{k})$.
This would yield the $z$ -transform of the moments and later we show how we switch back to $N$-space.
We now turn to finding  $G(z,\mathbf{k})$ in the case of obstacles.

The general case \textit{with} obstacles can be obtained by relying on the scattering formalism borrowed from quantum mechanics \cite{ballentine2014quantum}.
The dynamics in the presence of randomly distributed obstacles on the lattice is generated by the modified single jump propagator $\hat{M} = \hat{M}_0 + \hat{V}$, where $\hat{V}$ cancels the transitions from and to the obstacles. 
Furthermore, in our calculations we allow the tracer to start at an obstacle site, which then remains immobile.
The potential $\hat{v}_1\left(\mathbf{s}_1\right)$ for a single obstacle at site $\mathbf{s}_1$ cancels the transition probabilities from and into the obstacle position.
Therefore the only non vanishing elements of the matrix $\hat{v}_1\left(\mathbf{s}_1\right)$ involve the obstacle site and its immediate four neighbors.
In Sec.~\ref{subsection:scattering_matrix} we will explicitly show the resulting matrix $\hat{v}_1\left(\mathbf{s}_1\right)$. 
Now for $J$ obstacles (or impurities) the total potential will be $\hat{V} = \sum_{i=1}^{J} \hat{v}_i$.
We define the obstacle density as
\begin{equation} \label{eq:n_def}
n = J / L^2.
\end{equation}
Strictly speaking this definition does not properly account for obstacles that could occur as neighbors since then they will affect each others potential. 
But in the limit of large lattices $L\to \infty$ and small densities $n$ such realizations rarely occur and therefore we neglect them in our work. 
The propagator in the presence of a fixed obstacle realization $\hat{V} = \sum_{i=1}^{J} \hat{v}_i$  is related to the free propagator $\hat{G}_0$ via a Lippmann-Schwinger equation \cite{ballentine2014quantum}
%%%%%%%%%%%%%%%%%%%%%%%%%%%%%%%%%
\begin{equation}
    \hat{G} = \hat{G}_0 + \hat{G}_0 \hat{V} \hat{G}.
\end{equation}
%%%%%%%%%%%%%%%%%%%%%%%%%%%%%%%%%%
By iterating we can express $\hat{G}$ as
%%%%%%%%%%%%%%%%%%
\begin{equation}
\begin{aligned}
    \hat{G} &= \hat{G}_0 + \hat{G}_0 \hat{V} \left( \hat{G}_0 + \hat{G}_0 \hat{V} \hat{G} \right)\\
    &= \hat{G}_0 + \hat{G}_0 \hat{V} \hat{G}_0 + \hat{G}_0 \hat{V} \hat{G}_0 \hat{V} \hat{G} ,
\end{aligned}
\end{equation}
%%%%%%%%%%%%%%%%%%%%%%
and this can be further expended.
By following this line of reasoning, the propagator can be expressed as follows
\begin{equation} \label{eq:G_T_matrix}
\hat{G} = \hat{G}_0 + \hat{G}_0 \hat{T} \hat{G}_0 ,
\end{equation}
with the scattering matrix $\hat{T}$ defined by
%%%%%%%%%%%%%%%%%%%%%%%%%%%%%
\begin{eqnarray}
\hat{T} &=&\hat{V}+\hat{V} \hat{G}_0 \hat{V}+\hat{V} \hat{G}_0 \hat{V} \hat{G}_0 \hat{V}+\cdots \\ \nonumber
&=&\sum_{i=1}^{J} \hat{v}_i+\sum_{j, k=1}^{J} \hat{v}_j \hat{G}_0 \hat{v}_k+\sum_{l, m, n=1}^{J} \hat{v}_l \hat{G}_0 \hat{v}_m \hat{G}_0 \hat{v}_n+\cdots.
\end{eqnarray}
%%%%%%%%%%%%%%%%%%%%%%%%%%
We use the single-obstacle scattering operator $\hat{t}_i$ which encodes all possible collisions with obstacle $\hat{v}_i$ defined in $z$-space by
%%%%%%%%%%%%%%%%%%%%%%%%%%%%
\begin{equation}
\hat{t}_i=\hat{v}_i+\hat{v}_i \hat{G}_0 \hat{v}_i+\hat{v}_i \hat{G}_0 \hat{v}_i \hat{G}_0 \hat{v}_i+\cdots,
\end{equation}
%%%%%%%%%%%%%%%%%%%%%%%%5
and express $\hat{T}$ through a multiple scattering expansion, which reads
%%%%%%%%%%%%%%%%%%%%%%
\begin{equation}
\hat{T}=\sum_{i=1}^{J} \hat{t}_i+\sum_{\substack{j, k=1 \\ j \neq k}}^{J} \hat{t}_j \hat{G}_0 \hat{t}_k+\sum_{\substack{l, m, n=1 \\ l \neq m, m \neq n}}^{J} \hat{t}_l \hat{G}_0 \hat{t}_m \hat{G}_0 \hat{t}_n+\cdots.
\end{equation}
%%%%%%%%%%%%%%%%%%%%%%%%%%%%
We are not interested in the dynamics for individual realizations of the obstacle configurations but only in disorder-averaged properties. We use $[\cdot]_{\text{av}}$ to indicate an average over all realizations of the disorder. In particular, after disorder-averaging the system is translationally invariant. The convenient starting point for disorder averaging is Eq.~\eqref{eq:G_T_matrix}. In particular, 
$[\hat{T}]_{\text{av}}$ is translationally invariant in space and thus diagonal in the plane-wave basis.
Contributions in first order of the density can now be identified with the forward-scattering amplitude $t_i(z, \mathbf{k})=\langle\mathbf{k}|\hat{t}_i| \mathbf{k}\rangle$ of a single obstacle placed in a random location on the lattice, therefore
%%%%%%%%%%%%%%%%
\begin{equation} \label{eq:T}
\left\langle\mathbf{k}\left|[\hat{T}]_{\mathrm{\text{av}}}\right| \mathbf{k}\right\rangle = n L^2 t_i(z, \mathbf{k})+O\left(n^2\right).
\end{equation}
%%%%%%%%%%%%%%%%%%%%
Eq.~\eqref{eq:T} states that the disorder averaged scattering operator is proportional, in $\mathbf{k}$-space, to the forward scattering amplitude of a single obstacle.
The Lippmann-Schwinger equation \cite{ballentine2014quantum} allows us to express $\hat{t}_i$ as
%%%%%%%%%%%%%
\begin{equation}
\hat{t}_i=\hat{v}_i+\hat{v}_i \hat{G}_0 \hat{t}_i=\hat{v}_i+\hat{t}_i \hat{G}_0 \hat{v}_i,
\end{equation}
%%%%%%%%%%%%%%%%%%
therefore
%%%%%%%%%%%%%%%%%%%%%%
\begin{eqnarray} \label{eq:t_real-space}
    \left\langle\mathbf{r}|\hat{t}_i| \mathbf{r}^{\prime}\right\rangle&=&\left\langle\mathbf{r}\left|\hat{v}_i\left(1-\hat{G}_0 \hat{v}_i\right)^{-1}\right| \mathbf{r}^{\prime}\right\rangle \nonumber \\
    &=&\left\langle\mathbf{r}\left|\left(1-\hat{v}_i \hat{G}_0\right)^{-1} \hat{v}_i\right| \mathbf{r}^{\prime}\right\rangle.
\end{eqnarray}
%%%%%%%%%%%%%%%%%%%%%%%%%%5
Since $\hat{v}_i$ only has nonvanishing contributions for the obstacle site and its nearest neighbors, the calculation of  $\left\langle\mathbf{r}|\hat{t}_i| \mathbf{r}^{\prime}\right\rangle$ in Eq. (\ref{eq:t_real-space}) reduces to a $5 \times 5$  matrix inversion problem.

The real-space matrix elements $\langle\mathbf{r}|\hat{G}_0| \mathbf{r}^{\prime}\rangle$ of the free propagator for sites around the obstacle are expressed in terms of complete elliptic integrals of the first and second kind (we show this in Sec.~ \ref{subsection:matrix_elements_of_G}).
The forward-scattering amplitude is then obtained by a change of basis with,
%%%%%%%%%%%%%%%%%%%%%%%%%%%%%%
\begin{equation} \label{eq:t_k-space}
L^2 t_i(z, \mathbf{k})=\sum_{\mathbf{r}, \mathbf{r}^{\prime}} e^{\mathrm{i} \mathbf{k} \cdot\left(\mathbf{r}-\mathbf{r}^{\prime}\right)}\left\langle\mathbf{r}|\hat{t}_i| \mathbf{r}^{\prime}\right\rangle ,
\end{equation}
%%%%%%%%%%%%%%%%%%%%%%%%%%%%
where the sum effectively extends only over the obstacle site and its nearest neighbors.
The disorder-averaged propagator to first order in the density of obstacles $n$ (defined in Eq.~(\ref{eq:n_def})) is then according to Eqs.~(\ref{eq:G_T_matrix}) and~(\ref{eq:T}) 
%%%%%%%%%%%%%%%%%%%%%
\begin{equation} \label{eq:Green_k}
[G]_{\mathrm{\text{av}}}(z, \mathbf{k})=G_0(z, \mathbf{k})+n L^2 G_0(z, \mathbf{k})^2 t_i(z, \mathbf{k})+O\left(n^2\right).
\end{equation}
%%%%%%%%%%%%%%%%%%%%
We still need to account for one more correction.
Since the starting position is random, the tracer can start at an obstacle site where it will stay immobile forever. This obstacle can be surrounded by additional obstacle sites but to first order in the density this can be ignored by assuming obstacle sites are far enough from each other.
Starting movement at an obstacle site is not physical, therefore we correct for this behavior by simply multiplying the propagator $[G]_{\mathrm{\text{av}}}(z, \mathbf{k})$ with $1 / (1-n) = 1 + n +O(n^2)$, where $(1 - n)$ is the fraction of free lattice sites. We denote the corrected propagator by ${[G]}_{\mathrm{\text{av}}}^c(z, \mathbf{k})$ and keep only terms to first order in $n$
%%%%%%%%%%%%%%%%%%%%%
\begin{eqnarray} \label{eq:Green_k_corrected}
[G]_{\mathrm{\text{av}}}^c(z, \mathbf{k}) &=& G_0(z, \mathbf{k})+n \big[ G_0(z, \mathbf{k}) \\ \nonumber
& & + L^2 G_0(z, \mathbf{k})^2 t_i(z, \mathbf{k}) \big] +O\left(n^2\right).
\end{eqnarray}
%%%%%%%%%%%%%%%%%%%%

\section{The propagator} \label{section:the_propagator}

As shown in Sec. \ref{sec:the_model}, finding the propagator in the case of obstacles in $\mathbf{k}$-space will enable us to calculate the moments of displacement.
According to Eq. (\ref{eq:Green_k}), the missing piece is the single obstacle forward-scattering amplitude $t_i(z, \mathbf{k})=\langle\mathbf{k}|\hat{t}_i| \mathbf{k}\rangle$.
Since $\hat{t}_i$ is provided in Eq. (\ref{eq:t_real-space}) in real-space, we have to find the ingredients $\langle \mathbf{r} | \hat{G}_0 | \mathbf{r}' \rangle$ and $\langle \mathbf{r} | \hat{v}_i | \mathbf{r}' \rangle$ and then transition to $\mathbf{k}$-space.
%%%%%%%%%%%%%%%%%%%%%%%%%%%%%%%

\subsection{The single obstacle potential matrix $\hat{v}_i$} \label{subsection:single_potential_matrix}

For a reflective obstacle placed at $\mathbf{s}_i$ the new transition probabilities ($\hat{M} + \hat{v}_i$) should cancel the transition from and to the obstacle site. 
Hence $\hat{v}_i$ is a $5 \times 5$ matrix with elements $\langle \mathbf{r} | \hat{v}_i | \mathbf{r}' \rangle$ that correspond to the transition from $\mathbf{r}'$ to $\mathbf{r}$ with the basis  $\mathbf{r},\mathbf{r}^{\prime} \in \left\{\mathbf{s}_i \pm \mathbf{e}_y, \mathbf{s}_i \pm \mathbf{e}_x, \mathbf{s}_i \right\}$ and we set $\mathbf{s}_i$ for convenience to be zero.
The column index is $\mathbf{r}'$ and the row index is $\mathbf{r}$.
The reflective potential cancels the transition from and to the obstacle site, hence $\left\langle\mathbf{s}_i\left|\hat{v}_i\right| \mathbf{s}_i-\mathbf{d}\right\rangle=-W(\mathbf{d})$ and $\left\langle\mathbf{s}_i-\mathbf{d}\left|\hat{v}_i\right| \mathbf{s}_i\right\rangle=-W(-\mathbf{d})$.
If the tracer is at a neighboring site and attempts to jump to $\mathbf{s}_i$ it is scattered back to its original position, therefore $\left\langle\mathbf{s}_i-\mathbf{d}\left|\hat{v}_i\right| \mathbf{s}_i-\mathbf{d}\right\rangle=W(\mathbf{d})$.
Finally if the tracer is at the obstacle site it is stuck there forever, $\left\langle\mathbf{s}_i\left|\hat{v}_i\right| \mathbf{s}_i\right\rangle=1$.
We order the rows and columns for the matrix forms of the operators in the real-space basis via the scheme 
\begin{equation}
\begin{array}{lll} 
& 1 & \\
2 & 3 & 4 \\
& 5 &
\end{array}
\end{equation}
where the obstacle site is located at the origin $\mathbf{0}$ numbered by 3.
Consequently the order is $\mathbf{e}_y$, $-\mathbf{e}_x$, $\mathbf{0}$, $\mathbf{e}_x$, $-\mathbf{e}_y$.
The resulting potential matrix in our basis is now
%%%%%%%%%%%%%%%%%%%%%%%%%%%%%%%%%
\begin{equation} \label{eq:potential_matrix}
\begin{aligned}
&
\hat{v}_i =\Gamma\left(\begin{array}{ccccc}
1 & 0 & -1 & 0 & 0 \\
0 & e^{F / 2} & -e^{-F / 2} & 0 & 0 \\
-1 & -e^{F / 2} & 1 / \Gamma & -e^{-F / 2} & -1 \\
0 & 0 & -e^{F / 2} & e^{-F / 2} & 0 \\
0 & 0 & -1 & 0 & 1
\end{array}\right).
\\&
\,
\end{aligned}
\end{equation}
%%%%%%%%%%%%%%%%%%%%%%%%%%%%%%%%%
\subsection{The matrix elements of $\hat{G}_0(z)$} \label{subsection:matrix_elements_of_G}
The obstacle-free propagator is provided in Eq. (\ref{eq:G0_k-space}) in $\mathbf{k}$-space, to find the matrix elements in $\mathbf{r}$-space we perform an inversion to find
\begin{equation} \label{eq:z-transform_definition}
    \langle \mathbf{r} | \hat{G}_0(z) | \mathbf{r'} \rangle =  \int_{-\pi}^\pi \int_{-\pi}^\pi  \frac{d \mathbf{k}}{(2 \pi)^2}\frac{\exp \left(-i \mathbf{k} \cdot\left(\mathbf{r}-\mathbf{r}^{\prime} \right)\right) }{1-z\lambda(\mathbf{k})}.
\end{equation}

The entries of the matrix $\hat{G}_0$ are the same as defined in Sec.~\ref{subsection:single_potential_matrix}, with
$\mathbf{r},\mathbf{r}^{\prime} =  \mathbf{e}_y, -\mathbf{e}_x, 0, \mathbf{e}_x,-\mathbf{e}_y $.
The eigenvalues $\lambda(\mathbf{k})$ of the single jump matrix $\hat{M}_0$ follow directly from translational invariance in the $\mathbf{k}$-basis
%%%%%%%%%%%%%%%%%%%%%%%%%%%%%%%%%%%%%
\begin{equation}
    \langle \mathbf{k} | \hat{M}_0 | \mathbf{k} \rangle =
    \sum_{\mathbf{d} \in \mathcal{N}} \exp[i \mathbf{k} \cdot \mathbf{d}] W(\mathbf{d})
\end{equation}
%%%%%%%%%%%%%%%%%%%%%%%%%%%%%%%%%%%%%%%%%
with the result
%%%%%%%%%%%%%%%%%%%%%%5
\begin{eqnarray} \label{eq:lambda_biased}
\lambda(\mathbf{k}) &=& 2 \Gamma \Big[ \cos (k_x) \cosh (F/2)   \nonumber
\\
     & & + i \sin (k_x) \sinh (F/2) + \cos (k_y) \Big],
\end{eqnarray}
%%%%%%%%%%%%%%%%%%%%%%%%
$\Gamma$ was defined in Eq. (\ref{eq:transition_normalization}) to be the normalization of the transition probabilities.
We show in the Appendix~\ref{sec:the_matrix_elements} that the matrix elements of the free propagator can also be expressed in the following form
%%%%%%%%%%%%%%%%%%%%%%
\begin{eqnarray} \label{eq:z-transform_general_solution}
    \langle \mathbf{r} | \hat{G}_0(z) | \mathbf{r'} \rangle &=& e^{F (x - x') / 2} \int_0^{\infty} d s e^{-s} \\
    & & \times   I_{|x-x'|}\left(2 \Gamma s z \right) I_{|y-y'|}\left(2 \Gamma s z\right) \nonumber,
\end{eqnarray}
%%%%%%%%%%%%%%%%%%%%%%
where $I_m(\dots)$ is a modified Bessel function of the first kind of integer
order $m$. 
Here, the exponential outside of the integral accounts for the asymmetry introduced by the bias and the remaining terms are the solution of the symmetrical problem but with the diffusion coefficient set to be $\Gamma(F)$ instead of $\Gamma(F=0)$. 
As such we denote the obstacle-free unbiased part by
%%%%%%%%%%%%%%%%%
\begin{equation} \label{eq:g_formula}
        g_{x-x',y-y'} = \int_0^{\infty} d s\, e^{-s} I_{|x-x'|}\left(2 \Gamma s z \right) I_{|y-y'|}\left(2 \Gamma s z\right),
\end{equation}
%%%%%%%%%%%%%%%%%%%
and we obtain $\langle \mathbf{r} | \hat{G}_0(z) | \mathbf{r'} \rangle = e^{F (x - x') / 2} g_{x-x',y-y'}$. By using the symmetry of the unbiased part $g_{x y}$, with respect to $x$, $y$ and also the fact that $g_{x y}=g_{y x}$ we obtain for the matrix $\hat{G}_0 (z)$ the following representation 
\begin{widetext}
\begin{equation} \label{eq:G0_matrix_form}
\hat{G}_0(z)=\left(\begin{array}{ccccc}
g_{00} & e^{F / 2} g_{11} & g_{10} & e^{-F / 2} g_{11} & g_{20} \\
e^{-F / 2} g_{11} & g_{00} & e^{-F / 2} g_{10} & e^{-F} g_{20} & e^{-F / 2} g_{11} \\
g_{10} & e^{F / 2} g_{10} & g_{00} & e^{-F / 2} g_{10} & g_{10} \\
e^{F / 2} g_{11} & e^F g_{20} & e^{F / 2} g_{10} & g_{00} & e^{F / 2} g_{11} \\
g_{20} & e^{F / 2} g_{11} & g_{10} & e^{-F / 2} g_{11} & g_{00}
\end{array}\right).
\end{equation}
\end{widetext}
We note that by virtue of the underlying dihedral symmetry~\cite{brummelhuis1988single} that only four elementary propagators are needed to be explicitly calculated, they are: $g_{00}$,  $g_{10}$, $g_{20}$ and $g_{11}$.
As illustrated in Appendix~\ref{sec:force_free_prop_relations} these four propagators satisfy a set of linear relationship that eventyally reduce the number of independent propagators from four to two.
The resulting relations are
\begin{align} \label{eq:symmetrical_connections}    
        g_{00} &= 1 + 4 \Gamma z g_{10}, \nonumber \\
        g_{10} &= \Gamma z ( g_{00} + g_{20} + 2 g_{11} ).
\end{align}
We turn now to evaluating $g_{00}$ and $g_{11}$.
From Eq.~\eqref{eq:g_formula},
\begin{equation} \label{eq:g00}
    g_{00} = \int_0^{\infty} d s e^{-s} I_0\left(2 \Gamma s z\right)^2 = \frac{2}{\pi} \boldsymbol{K}(16 \Gamma ^2 z^2)
\end{equation}
\begin{equation} \label{eq:g11}
\begin{aligned}
    g_{11} &= \int_0^{\infty} d s e^{-s} I_1\left(2 \Gamma s z\right)^2\\
    &= \frac{2}{\pi (4 \Gamma z)^2}\Big\{\left[2-(4 \Gamma z)^2\right] \boldsymbol{K}\left[(4 \Gamma z)^2\right]\\
    &-2 \boldsymbol{E}\left[(4 \Gamma z)^2\right]\Big\}
\end{aligned}
\end{equation}
where
%%%%%%%%%%%%%%%%%%55
\begin{equation}
\boldsymbol{K}(k) = \int_0^{\pi / 2} \frac{d \alpha}{\sqrt{1-k \sin ^2 \alpha}}
\end{equation}
%%%%%%%%%%%%%%%%%%%%55
and
\begin{equation}
    \boldsymbol{E}(k) = \int_0^{\pi /2} \sqrt{1-k \sin ^2 \alpha} \; d \alpha
\end{equation}
are the complete elliptic integral of the first and second kind respectively.
From Eq.~\eqref{eq:symmetrical_connections}, $g_{10}$ and $g_{20}$ are easily found.

\subsection{Scattering matrix $\hat{t}_i$} \label{subsection:scattering_matrix}
To sum up the work done so far; we have found all the ingredients required to calculate the single-obstacle scattering matrix $\hat{t}_i$ in $z$-space and in the plane-wave basis. By using 
Eq.\eqref{eq:t_k-space} together with Eq.\eqref{eq:t_real-space}, we find
\begin{equation} \label{eq:t_formula}
\begin{aligned}
    L^2 t_i(z, \mathbf{k})&=L^2\langle\mathbf{k}|\hat{t}_i| \mathbf{k}\rangle=\\
    &\sum_{\mathbf{r}, \mathbf{r}^{\prime}} \exp \left[i \mathbf{k} \cdot\left(\mathbf{r}-\mathbf{r}^{\prime}\right)\right]\left\langle\mathbf{r}\left|\hat{v}_i\left(1-\hat{G}_0 \hat{v}_i\right)^{-1}\right| \mathbf{r}^{\prime}\right\rangle.
\end{aligned}
\end{equation}
where $\hat{v}_i$ is given in Eq.\eqref{eq:potential_matrix} and $\hat{G}_0$ in Eq.~\eqref{eq:G0_matrix_form} is a function of the four obstacle-free propagators, $g_{00}, \quad g_{10}, \quad g_{20}, \quad g_{11} $. The problem turns into a $5 \times 5$ matrix inversion problem which we solve using computer algebra. 
Finding $t_i(z, \mathbf{k})$ allows us to obtain the propagator ${[G]}_{\mathrm{\text{av}}}^c(z, \mathbf{k})$ for the system with obstacles using Eq.~\eqref{eq:Green_k_corrected}. As mentioned before, this propagator is also the moment generating function. Thus, we derive the moments of displacement by taking the appropriate derivative of $[G]_{\mathrm{\text{av}}}^c(z, \mathbf{k})$.
The $m$-th moment of displacement along the $x$ axis, i.e., in parallel to the direction of the applied force (see Fig.~\ref{Fig:1}), is now
\begin{align} \label{eq:moments_formula}
    \langle \tilde{x}^m(z) \rangle & := \sum_{N=0}^\infty \langle x(N)^m \rangle z^N \nonumber \\
&=  (-i)^{m}\partial^{m} [G]_{\text{av}}^c(z, \mathbf{k}) / \partial k_x^{m} |_{\mathbf{k=0}}.
\end{align}
Note that in the $z$ domain, we need to distinguish $\tilde{x}^m(z)$ from $\tilde{x}(z)^m$.
We used the notation $\langle \cdots \rangle$ in order to indicate the average over the randomness (many obstacle realizations and different trajectories).
In what follows, we investigate the behavior of these moments for any number of steps $N$, and in particular the convergence towards the terminal velocity $v_\infty := \lim_{N\to \infty}\langle x(N) \rangle / N$ and the variance $\left[ \langle x(N)^2 \rangle - \langle x(N) \rangle ^2 \right]$ as a function of $N$. The moments are derived for arbitrary force $F$ correct to first order in the density of obstacles $n$ and compared to computer simulations to find the range of validity.

\section{First Moment}
Equation ~\eqref{eq:moments_formula} allows computing the first moment of the displacement for any $z$ and force $F$,
\begin{align} \label{eq:first_moment_in_z-space_definition}
    \langle \tilde{x} (z) \rangle =& -i \frac{\partial}{\partial k_x} 
      \Big[ (1 + n) G_0 (z, \mathbf{k}) \\ \nonumber
      &+ n L^2 G_0 (z, \mathbf{k})^2 t_i(z, \mathbf{k})\Big] \Bigg|_{\mathbf{k}=0}.
\end{align}
We remind the reader that $G_0 (z, \mathbf{k})$ is the obstacle-free propagator and is provided by Eq. ~\eqref{eq:G0_k-space} giving $G_0 (z, \mathbf{k}=0) = 1/(1-z)$. 
By expanding Eq.~\eqref{eq:first_moment_in_z-space_definition} we obtain 
\begin{align}
    \langle \tilde{x} (z) \rangle =& \left. (1+n) \left(-i \frac{\partial}{\partial k_x} G_0(z,\mathbf{k}) \right) \right|_{\mathbf{k}=0} \\ \nonumber
    &  + n \left[ -2iL^2 G_0(z,\mathbf{k}) t_i(z,\mathbf{k}) \left( \frac{\partial G_0}{\partial k_x} \right) \right.\\ \nonumber
    &  +\left. \left. G_0(z,\mathbf{k})^2 \left( -iL^2 \frac{\partial t_i(\mathbf{k})}{\partial k_x} \right) \right] \right|_{\mathbf{k}=0}.
\end{align}
Additionally, by symmetry of the problem and verified using computer algebra, we find $t_i(z,\mathbf{k}=0)=0$.
We now have an expression for $\langle \tilde{x} (z) \rangle$,
%%%%%%%%%%%%%%%%%%%%%%%%%%%%%%%
\begin{eqnarray} \label{eq:x(z)}
    \langle \tilde{x} (z) \rangle &-& (1 + n) \langle \tilde{x}_0(z) \rangle = \nonumber \\
    &=& n  \frac{1}{(1-z)^2} \left.\left(-\mathrm{i} L^2 \frac{\partial t_i(z, \mathbf{k})}{\partial k_x} \right) \right|_{\mathbf{k}=0},
\end{eqnarray}
%%%%%%%%%%%%%%%%%%%%%%%%%%%%%%%
we find it using computer algebra. The resulting expression is extremely long and provided in the appendix [Eq.~\eqref{eq:full_first_moment}].
Here $\langle \tilde{x}_0(z) \rangle$ is the solution of the bare dynamics which we know after transitioning to $N$-space must be $\langle x_0(N) \rangle = \tanh (F/4) N$. This is found by averaging over the displacement of a single step (see Sec.~\ref{sec:the_model}) and multiplying by $N$.
As the solution in Eq.~\eqref{eq:x(z)} is in $z$-space, we look for the solution for any number of steps $N$ by inverting the process.
Inverting the $z$-transform in Eq.~\eqref{eq:moments_formula}, we obtain the average displacement
\begin{equation} \label{eq:inverse_z_transform}
    \langle x(N) \rangle = \frac{1}{N!} \left. \frac{d^N}{dz^N} \langle \tilde{x} (z) \rangle \right|_{z=0}.
\end{equation}
This provides the result for any $N$ and $F$.
Successively, we will derive in Sec.~\ref{section:terminal_velocity} the terminal velocity and we will use this formula for $\langle x(N) \rangle$ in order to test the convergence rate towards the terminal velocity in theory and simulations.

\section{Terminal velocity} \label{section:terminal_velocity}
To determine the asymptotic behavior of the discrete velocity, $v_{\infty} := \lim_{N\to \infty} \left( \langle x(N) \rangle / N \right)$, we make use of the Tauberian theorem~\cite{klafter2011first, weiss1994aspects, hughes1995random} (see the appendix for the exact method).
The Tauberian Theorem allows transitioning to $N$-space and to find $\langle x(N) \rangle$ for large $N$ from the behavior of its generating function $\langle \tilde{x} (z) \rangle =\sum_{N=0}^{\infty} \langle x(N) \rangle z^N$ when $z\to 1$.
We use in Eq.~\eqref{eq:x(z)} the asymptotic relation, $\left.\left(-\mathrm{i} L^2  \partial t_i(z, \mathbf{k})/\partial k_x \right)\right|_{\mathbf{k}=0} = \Delta v_{\infty}+O(1-z)$ obtained in Appendix ~\ref{section:asymptotic_limits} where $\Delta v_{\infty}$ is a function of  $F$ only.
Consequently the $1/(1-z)^2$ term in Eq.~\eqref{eq:x(z)} transforms to a linear growth in $N$ and we obtain,
%%%%%%%%%%%%%%%%%%%%%%%%%%%%%%%
\begin{eqnarray} \label{eq:x(N)}
    \langle x(N) \rangle \sim (1+n) v_0 N + n \Delta v_{\infty} N,
\end{eqnarray}
%%%%%%%%%%%%%%%%%%%%%%%%%%%%%%%
as $N\to \infty$. Here  $v_0=\tanh (F/4)$ denotes the velocity of the obstacle-free lattice. Therefore,
\begin{eqnarray} \label{eq:discrete_terminal_velocity}
    v_{\infty} &=& \lim_{N \to \infty} (\langle x(N) \rangle) / N \nonumber \\
    &=& v_0 + n (v_0 +\Delta v_{\infty}).
\end{eqnarray}
% %%%%%%%%%%%%%%%%%%%%%%%%%%%%
The terminal behavior is now found for arbitrarily strong driving $F$ and small densities $n$, while the complete expression is very long [see Eq.~\eqref{eq:full_first_moment}], we can elaborate the behavior for small forces. Relying on computer algebra, we find
\begin{equation} \label{eq:x_avg_series}
    v_\infty = D_x F + \frac{n}{16} \left(\frac{\pi}{4} -1\right) F^3 \log (F) + O(F^3),
\end{equation}
which immediately highlights the nonanalytic behavior in $F$.
Here $D_x = [ 1 - n (\pi - 1) ] / 4$ corresponds to the diffusion coefficient in the $x$-direction in the absence of force as expected from linear response.
To make a connection with the continuous-time case, $N$ needs to be treated as a random variable, where the duration $\tau$ of a step is taken from a distribution $\psi(\tau)$. This is essentially the renewal theorem; the $m$-th moment in the time domain would now be given by  $\langle x(t)^m \rangle = \sum_N \langle x(N)^m \rangle Q_N(t)$,
where $Q_N(t)$ is the probability for having performed exactly $N$ steps until time $t$. For large times, this can be simplified, and the number of steps is switched with the total time passed $t$ divided by the expected time $\tau$ of a jump (as long as it is finite). When the times between steps is exponentially distributed with a mean of one, we immediately recover the expression already found in a previous work~\cite{leitmann2013nonlinear} by just switching $N$ with $t$ in Eq.~\eqref{eq:full_first_moment}, therefore validating our result.

\begin{figure}[!htb]
\centering
\includegraphics[width=0.45\textwidth]{"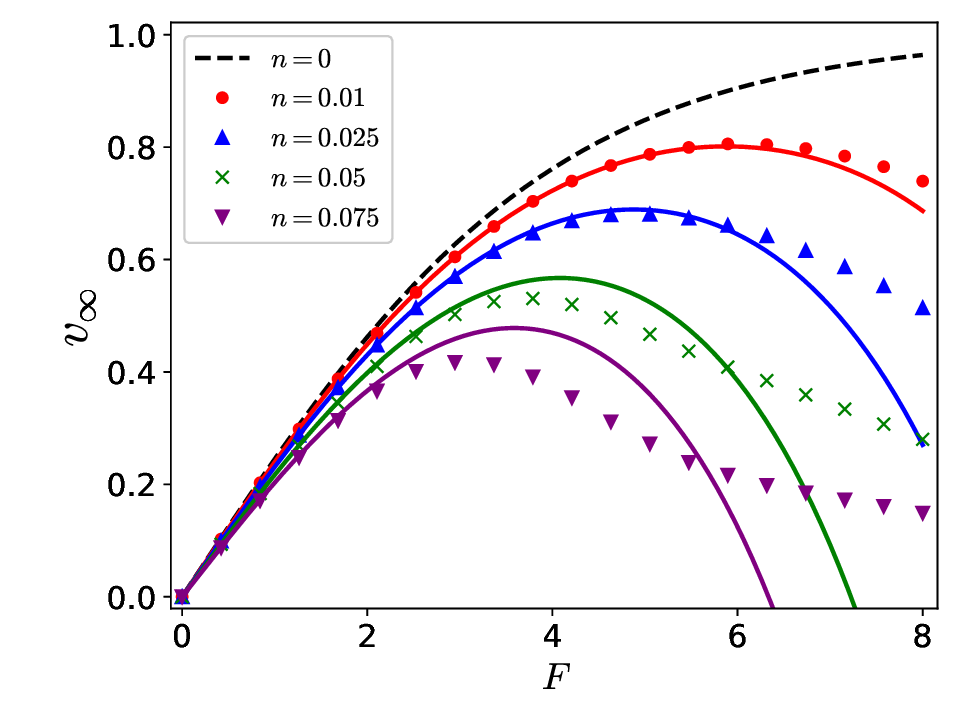"}
\caption{The terminal velocity $v_\infty := \lim_{N\to \infty} \langle x(N)\rangle /N $ of the tracer particle for different forces $F$ and obstacle densities $n$. Density increases from top to bottom. Solid lines correspond to the analytic solution and symbols represent simulation results. The dashed black line corresponds to the case of no obstacles.
}
\label{Fig:2}
\end{figure}

Our result in Eq.~\eqref{eq:x_avg_series} holds as long as obstacles are positioned far enough from each other such that sequences of collisions involving obstacles that have been encountered previously can be ignored.
Then, in our derivations, the assumption that $[T]_{\text{av}}(\mathbf{k})$, the disorder-averaged scattering operator in $\mathbf{k}$-space, is the sum of $nL^2$ identical single-obstacle forward-scattering amplitudes is correct [Eq.~\eqref{eq:T}].
To investigate the range of validity of the low-density expansion, we plot the terminal response as a function of $F$ in Fig.~\ref{Fig:2}. For each curve, there is a peak indicating a transition for stronger driving where the particle is more often stuck on an obstacle instead of going around it. Thus, the lattice Lorentz model displays negative differential mobility, a phenomenon that was seen in Ref.~\cite{benichou2014microscopic, basu2014mobility, jack2008negative, leitmann2018time, leitmann2013nonlinear} as well.
While the theoretical lines in Fig.~\ref{Fig:2} agree nicely with simulations, the suppression of the stationary velocity is underestimated at strong forces indicating that contributions of higher order in the density become relevant. In other words, the range of validity of our approach for the small obstacle densities is force-dependent.
%%%%%%%%%%%%%%%%%%%%%%%%%%%%%%%%%%%%%%%%%%%%%%%%%%%%%
% %%%%%%%%%%%%%%%%%%%%%%%%%%%%%
%\dan{The new part below about the convergence.}

\section{Approch  towards the terminal velocity} 
\label{sec:decay_rate}
\begin{figure}[!htb]
\centering
\includegraphics[width=0.45\textwidth]{"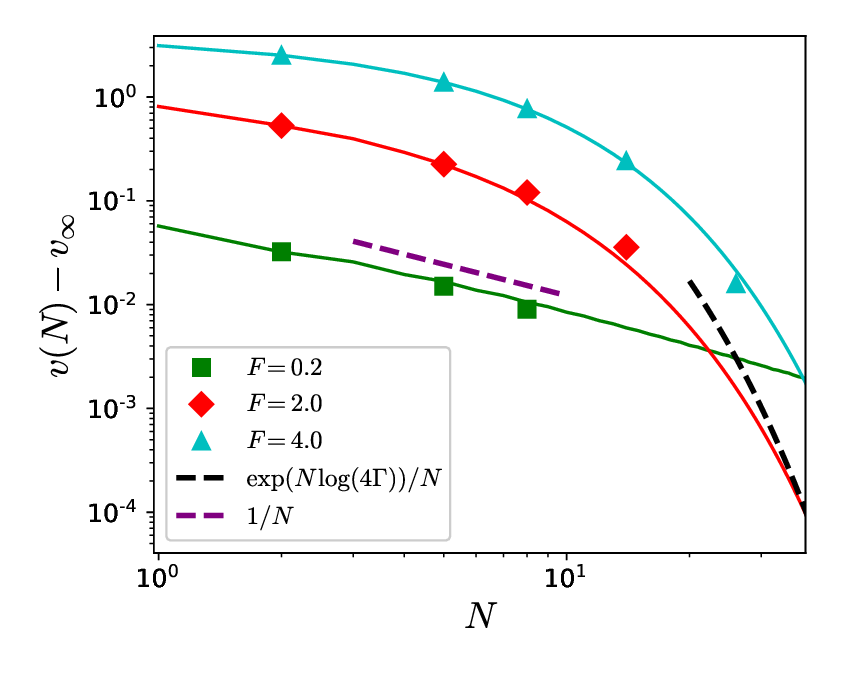"}
\caption{The approach of the instantaneous velocity $v(N)$ [Eq.~\eqref{eq:instant_v_def}] towards the discrete terminal velocity $v_\infty$ of the tracer particle for different forces $F$. Obstacle density is $n=0.001$. Force decreases from top to bottom. Solid lines correspond to the analytic solution in Eq.~\eqref{eq:inverse_z_transform} and Eq.~\eqref{eq:discrete_terminal_velocity} and symbols represent simulation results. The expected exponential behavior [Eq.\eqref{eq:exp_decay_proof}] is shown for reference in the black dashed line. The slope of linear response $1/N$ is also shown, in the purple dashed line, indicating the behavior for small $F$.}
\label{Fig:3}
\end{figure}

Next we investigate in more detail the instantaneous velocity defined by 
\begin{equation} \label{eq:instant_v_def}
v(N) := \langle x(N) - x(N-1) \rangle \qquad \text{for } N=1,\ldots
\end{equation}
where we set $v(0)  = 0$.  In the $z$-domain, it is directly related to the displacement via
\begin{equation}
\tilde{v}(z) = (1-z) \langle \tilde{x}(z) \rangle  
\end{equation}
The behavior of the instantaneous velocity at large step sizes is reflected in the poles of the $z$-transform. We have already identified a pole of order 2 in $\langle \tilde{x}(z) \rangle$ which translates to a simple pole in $\tilde{v}(z)$ reflecting that the velocity approaches a constant $v(N) \to v_\infty$ as $N\to \infty$. Further singular behavior emerges from the single-obstacle t-matrix $\hat{t}_i (z ,\mathbf{k})$ [Eq.~\eqref{eq:t_formula}]. Since the inversion of the $5\times 5$ matrix does not give rise to new singular behavior, all non-analytic properties are inherited by the matrix elements, i.e., by the 
\begin{equation} \label{eq:integrals_of_G}
    \langle \mathbf{r} | \hat{G}_0(z) | \mathbf{r'} \rangle =  \int_{-\pi}^\pi \int_{-\pi}^\pi  \frac{d \mathbf{k}}{(2 \pi)^2}\frac{\exp \left(-i \mathbf{k} \cdot\left(\mathbf{r}-\mathbf{r}^{\prime} \right)\right) }{1-z\lambda(\mathbf{k})}.
\end{equation}
For the case of $F=0$, there is a singularity at $z \to 1$ is originating from $\lambda(\mathbf{k} = \mathbf{0}) = 1$. 
%%%%%%%%%%%%%%%%%%%%%%%%%%%%%%%%%%%%%%%%%%%%%%%%%%%%%%%%%%%
In Sec.~\ref{subsection:matrix_elements_of_G} we have established a linear dependence of these integrals on the complete elliptic integrals of the first and second kinds, $\boldsymbol{K}$ and $\boldsymbol{E}$, respectively.
The leading term arises from the expansion of $\boldsymbol{K}$ around $z=1$,
\begin{equation} \label{eq:singularity_no_force}
    \left. \langle \mathbf{0} | \hat{G}_0(z) | \mathbf{0}  \rangle \right|_{F=0} = \frac{2}{\pi} \boldsymbol{K}(z^2) \sim -\frac{1}{\pi} \log(1-z),
\end{equation}
while $\boldsymbol{E}$ is subdominant (see Ref.~\cite{brummelhuis1988single}).  Upon Taylor expansion of the logarithm
\begin{eqnarray} \label{eq:elliptic_N_space}
- \log (1-z)=    \sum_{N=1}^{\infty} \frac{z^{N}}{N}, 
\end{eqnarray}
we read off that logarithm corresponds to an algebraic decay $\propto 1/N$ for $N\to \infty$. 
This corresponds to the decay rate expected from linear response when $F \to 0$ as seen in Fig.~\ref{Fig:3}. 
For the case of $F>0$, the integral in Eq.~\eqref{eq:integrals_of_G} is no longer divergent at $z=1$ but the singularity is shifted. 
 Using the same argument as in Eq.~\eqref{eq:singularity_no_force}, the leading term is found to 
\begin{equation} \label{eq:singularity_with_force}
     \langle \mathbf{0} | \hat{G}_0(z) | \mathbf{0}  \rangle = \frac{2}{\pi} \boldsymbol{K}(16 \Gamma ^2 z^2) \sim -\frac{1}{\pi} \log \left( 1-4 \Gamma z \right),
\end{equation}
i.e. we anticipate a singularity at $z=1/4 \Gamma > 1$. By the scaling property, the $z$-transform of $(4\Gamma)^{-N} [v(N)-v_\infty] $ is $\tilde{v}(z/4\Gamma)- v_\infty/(1- z / 4\Gamma )  \propto -\log(1-z)$ for $z\to 1$ , and we infer that the 
large-step behavior of the velocity in this case is 
\begin{eqnarray} \label{eq:exp_decay_proof}
    v(N) - v_\infty \propto \exp[ N \log (4 \Gamma) ] /N. 
\end{eqnarray}
Therefore the power law tail $\sim1/N$ no longer determines the asymptotic large-$N$ behavior for any finite bias $F$; rather the decay rate is exponentially fast  which can be elaborated further for small $F\ll 1$ to be $\sim \exp (-N F^2 / 16) / N$. A comparison of this analytical prediction with simulation is shown in  Fig.~\ref{Fig:3}. 
This result is consistent with the continuous case with exponential waiting times~\cite{leitmann2013nonlinear} where it was shown that the regular algebraic decay $\sim t^{-1}$ is elaborated to $\sim t^{-1} \exp (-F^2 t / 16)$ for $F \to 0$. 
The onset time of the exponential behavior can be found by comparing the decay rate with that of linear response, i.e., when $\exp[N \log( 4 \Gamma)] / N \ll 1/N$. We find $N \gg |1 /  \log(4 \Gamma)| = N_f$, where $N_f$ is the onset step-time. For small $F$, $N_f \sim 16/F^2 $, thus we establish that as the forces are smaller the onset time is delayed significantly as can be seen in Fig.~\ref{Fig:3}.

We remark that if we instead examine the average velocity, $\langle x(N) \rangle / N$, the exponential decay will become subdominant and we will observe the asymptotic behavior $\sim 1/N$ for $N \gg 1$.

% %%%%%%%%%%%%%%%%%%%%%%%%%%%%
% %%%%%%%%%%%%%%%%%%%%%%%%%%%%%

\section{Second moment}
By employing the same approach as for the first moment, the second moment of displacement in the $x$ direction, sometimes called mean-squared displacement (MSD) is found by taking the second derivative of the propagator in Eq.~\eqref{eq:Green_k_corrected}. From Eq.~\eqref{eq:moments_formula} we obtain,
\begin{align} \label{eq:x^2(Z)}
    \langle \tilde{x}^2(z) \rangle =& (-i)^2 \frac{\partial^2}{ \partial k_x^2} 
      \Big[ (1+n) G_0 (z, \mathbf{k}) \\ \nonumber
      &+ n L^2 G_0 (z, \mathbf{k})^2 t_i(z, \mathbf{k})\Big] \Bigg|_{\mathbf{k}=0}.
\end{align}
The first term here is the obstacle-free system which we denote by $\langle \tilde{x}_0^2(z) \rangle$. 
Using $G_0 (z, \mathbf{k}=0) = (1-z)^{-1}$ and that $t_i(z, \mathbf{k}=0)=0$, after some simplification we are left with
\begin{eqnarray} \label{eq:x(z)_squared}
    \langle \tilde{x}^2(z) \rangle &=& (1 + n) \langle \tilde{x}_0^2(z) \rangle - n \left[ 4 \left.\left( i L^2 \frac{\partial t_i (z, \mathbf{k})}{\partial k_x} \right)\right|_{\mathbf{k}=0} \right. \\ \nonumber
    & & \times \tanh \left(\frac{F}{4}\right) \frac{z}{(1-z)^3} \\ \nonumber
    & & + \left.\left. \frac{1}{(1-z)^2} \left( L^2 \frac{\partial^2}{\partial k_x^2} t_i (z, \mathbf{k}) \right)\right|_{\mathbf{k}=0} \right].
\end{eqnarray}
Making use of the inverse $z$-transform as in Eq.~\eqref{eq:inverse_z_transform}, we find the exact solution for $\langle x(N)^2 \rangle$ for any $N$ and $F$.
This result is used later to find the variance and how it behaves for small number of steps $N$ in Sec.~\ref{section:variance}. We continue to find the asymptotic behavior of the second moment in the limit of $N \to \infty$.

\section{Second moment  in the regime of large step numbers} \label{sec:second_moment_limit}
\begin{figure}[!htb]
\centering
\includegraphics[width=0.45\textwidth]{"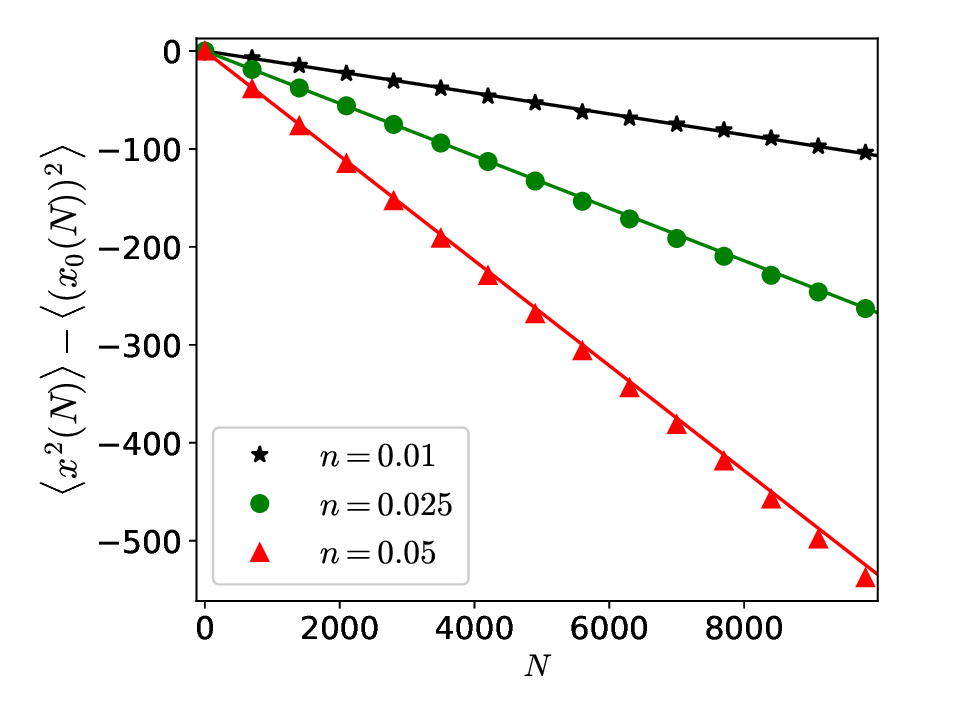"}
\caption{The first-order correction in $n$ to the second moment of the displacement when $F=0$. Solid lines correspond to the theoretical prediction in Eq.\eqref{eq:MSD}, symbols are simulation results.
Simulations are performed using the approach discussed in the App.~\ref{sec:simulations}.}
\label{Fig:4}
\end{figure}

We use the same methodology as in Sec.~\ref{section:terminal_velocity} to determine the asymptotic behavior of the MSD, $\langle x^2 (N) \rangle$, in the regime of $N\gg 1$.
We therefore use in Eq.~\eqref{eq:x(z)_squared} the $(1-z)$ expansions of the derivatives of the scattering matrix $t_i (z,\mathbf{k})$ up to order $O(1-z)$. 
Relying on the results derived in Appendix~\ref{section:asymptotic_limits}, we obtain $\left.\left(-\mathrm{i} L^2 \partial t_i(z, \mathbf{k}) / \partial k_x \right)\right|_{\mathbf{k}=0} = \Delta v_{\infty}+O(1-z)$ and 
$\left. \left( L^2 \partial^2 t_i(z,\mathbf{k}) / \partial^2 k_x \right) \right|_{\mathbf{k}=0}=-2 (v_0)^2 z/(1-z) +c_1(F) +O(1-z)$. Here, $c_1(F)$ here is very long expression that depends only on $F$, its explicit form can be inferred from Eq.~\eqref{eq:full_second_moment}.
Consequently, using again the Tauberian theorem, the $1/(1-z)^2$ term in Eq.~\eqref{eq:x(z)_squared} transform to $N$ and the $z/(1-z)^3$ term transforms to $N^2 / 2$.
Collecting results, we find asymptotically
\begin{align} \label{eq:MSD}
    \langle x(N)^2 \rangle \sim& (1+n) \langle x_0(N)^2 \rangle \nonumber \\
    &+n \left[ 2 v_0 \Delta v_{\infty} N^2 + v_0^2 N^2 + c_1(F) N \right],
\end{align}
for $N\to \infty$. Here the $\langle x_0 (N)^2 \rangle$ term is determined by the bare diffusion dynamics, $\langle x_0(N)^2 \rangle = \left( 1-\tanh ^2 (F/4) \right) N/2 + \tanh ^2 (F/4) N^2$.
This result is used later to determine the behavior of the variance in the regime $N\gg 1$ in Sec.~\ref{sec:diffusion_asymptotic}. For small forces, the expression in Eq.~\eqref{eq:MSD} can be simplified further, using $\Delta v_{\infty} = -\pi F / 4 +(\pi / 4 - 1) F^3 \log (F) / 16 + O(F^3)$ and approximating $c_1=-\pi/2 - (0.333005 +0.106403 \log (F)) F^2 + O(F^4)$, $c_1$ is given in full detail in the Appendix [Eq.~\eqref{eq:full_second_moment}], we obtain
\begin{eqnarray}
    &&\langle x(N)^2 \rangle - (1+n) \langle x_0(N)^2 \rangle \approx \\ \nonumber 
    &&-n \left\{ \frac{\left( 2 \pi - 1 \right)}{16} F^2 N^2 + \left[ \frac{\pi}{2} + \left(\frac{1}{3} +\frac{1}{10} \log (F) \right) F^2 \right] N\right\}.
\end{eqnarray}
The force-free term $[ 1-n(\pi -1) ] N / 2$ in the second moment of the displacement indicates that when no force is applied ($F=0$), the obstacles still obstruct the movement of the tracer.
This can be seen in the plot [see Fig.~\ref{Fig:4}] of the second moment of displacement in the $x$ direction. Higher-order corrections show that there is a complex non-linear dependence on the force, specifically the logarithmic dependence on $F$ that becomes even more relevant in the variance since the $N^2$ dependent terms drop, as we show in the next section.
%%%%%%%%%%%%%%%%%%%%%%%%%%%%
%%%%%%%%%%%%%%%%%%%%%%%%%%%%%

\section{The variance} \label{section:variance}
\begin{figure}[!htb]
\centering
\includegraphics[width=0.45\textwidth]{"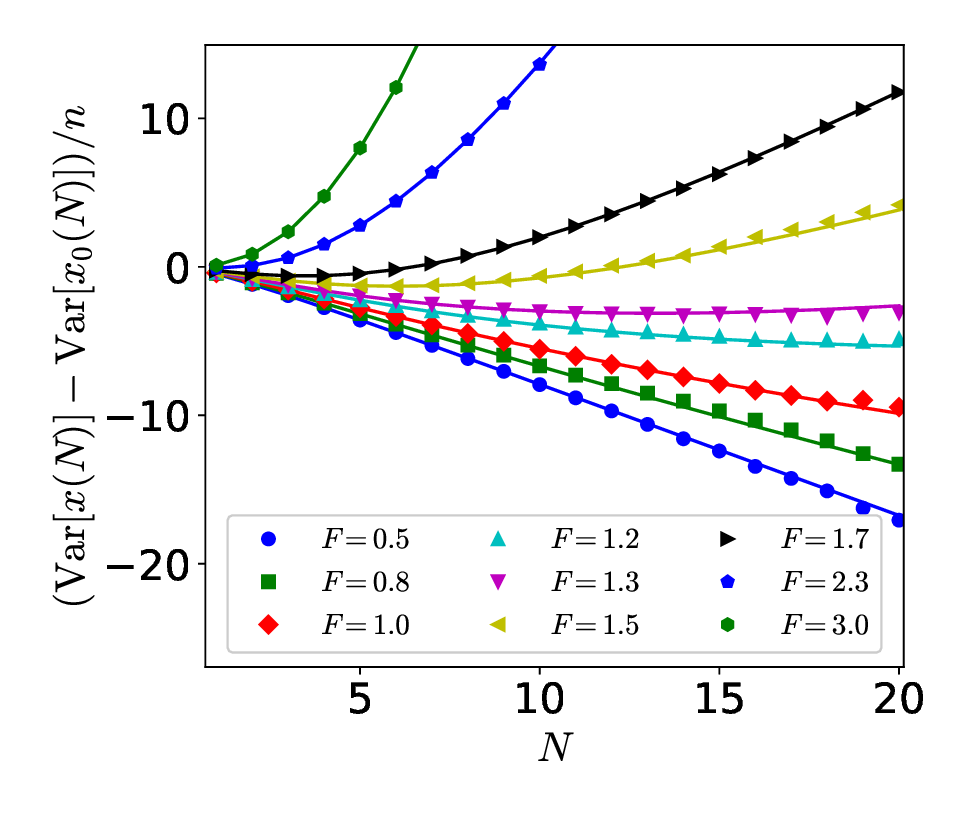"}
\caption{The first correction term in the obstacle density to the variance as a function of the number of steps. Force is increased from bottom to top. Solid lines correspond to the analytic solution and symbols represent simulation results. We see a transition from a negative contribution of the obstacle disorder to a positive one as the force is increased. This is accompanied with a transition from a linear dependence on steps $N$ to a power-law behavior. }
\label{Fig:5}
\end{figure}

\begin{figure}[!htb]
\centering
\includegraphics[width=0.45\textwidth]{"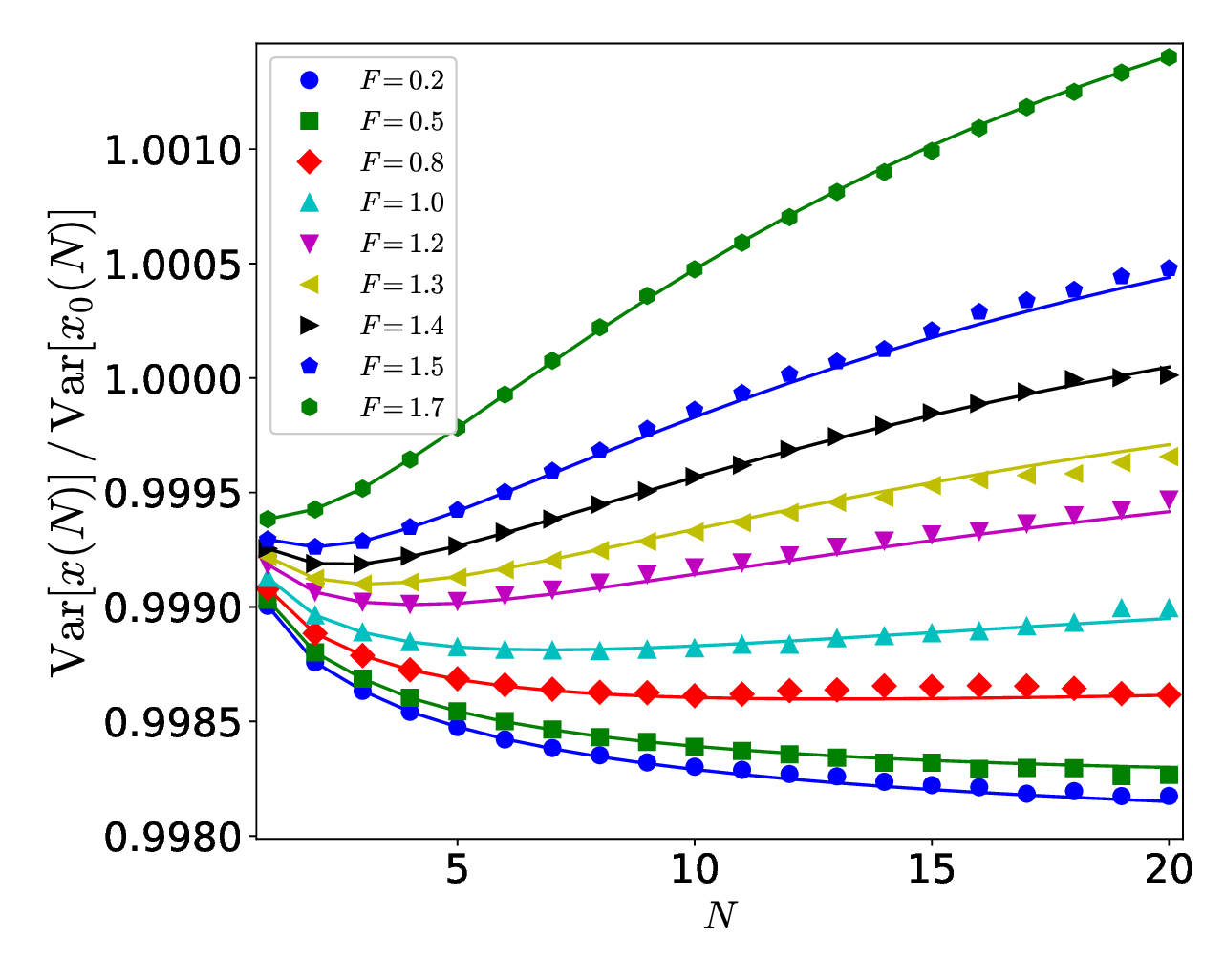"}
\caption{The behavior of the variance as a function of steps divided by the variance of the bare dynamics (no obstacles) for different forces. Force is increased from bottom to top. The obstacle density is $n=0.001$. Solid lines are the analytical solution for the inverse $z$-transform as explained in Sec.~\ref{section:variance} and the symbols are the simulations results. We see a transition from a negative contribution of the obstacle disorder to a positive one as the force is increased. }
\label{Fig:6}
\end{figure}

In order to determine the variance of the displacement $\textsf{Var}[x(N)]:=\langle x(N)^2 \rangle - \langle x(N) \rangle^2$, we remark that we take contributions only to first order in $O(n)$ when taking the square   of $\langle \tilde{x} (z) \rangle$ in Eq.~\eqref{eq:x(z)}. Then, the exact solution is determined by taking the inverse $z$-transform of $\langle \tilde{x}^2 (z) \rangle - \langle \tilde{x} (z) \rangle^2$ as was done in Eq.~\eqref{eq:inverse_z_transform} for the first moment.
In Fig.~\ref{Fig:5} and Fig.~\ref{Fig:6} we plot the variance in the direction of the force against numerical simulations for small number of steps $N$. 
For this purpose we use the variance of the bare dynamics $\textsf{Var}[x_0(N)] = \left( 1- \tanh^2 (F/4) \right) N / 2$.
We see in Fig.~\ref{Fig:6} that even for relatively small forces (in the range $1.0 \lesssim F \lesssim 1.7$), the behavior is non-monotonic in $N$ as the system is driven strongly out of equilibrium. An effect that does not appear at all for $F=0$ (compare Fig.~\ref{Fig:4}).
While non-monotonic behavior and breaking of linear response is expected for large forces, we see an additional effect. The system transitions from a negative contribution to the variance in the obstacle density to a positive one when the force is increased enough, $F \gtrsim 1.5$ [Fig.~\ref{Fig:5}]. Such a behavior has been seen in previous works for the lattice Lorentz model~\cite{illien2018nonequilibrium, leitmann2017time, illien2014velocity}, and we investigate it further in Sec.~\ref{sec:super_diffusion}. 

% %%%%%%%%%%%%%%%%%%%%%%%%%%%%
% %%%%%%%%%%%%%%%%%%%%%%%%%%%%%

% %%%%%%%%%%%%%%%%%%%%%%%%%%%%
% %%%%%%%%%%%%%%%%%%%%%%%%%%%%%

\section{Superdiffusion at intermediate steps} \label{sec:super_diffusion}
\begin{figure*}
\centering
\includegraphics[width=1.0\textwidth]{"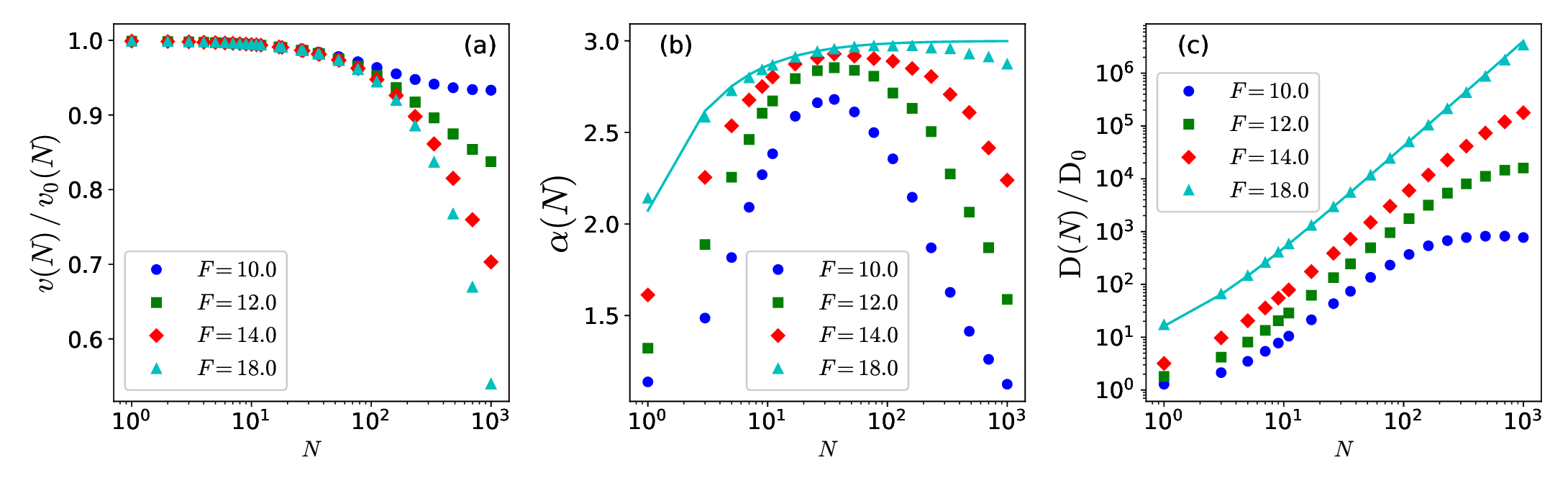"}
\caption{The discrete velocity $v(N) / v_0(N)$ [panel (a)], the local exponent  $\alpha (N)$ defined in Eq.~\eqref{eq:local_exponent} [panel (b)] and the diffusion coefficient $D(N)$ defined in Eq.~\eqref{eq:diffusion_definition} [panel (c)] for large increasing values of force $F$ with $n = 0.001$. In this regime the tracer mainly jumps only in the direction of the force. The analytic asymptotic solution ($F \to \infty$) of Eq.~\eqref{eq:var_superdiffusion} and Eq.~\eqref{eq:diffusion_asymptotic} are plotted as solid lines in panels (b) and (c) respectively.
}
\label{Fig:7}
\end{figure*}

The intuitive picture is that obstacles suppress the fluctuations in the direction of the force. Our results indicate that for forces large enough, increasing disorder leads to an enhancement.
The step-dependent behavior of the variance can be quantified in more detail by considering the step-dependent diffusion coefficient defined by a discrete step derivative [Fig.\ref{Fig:7}(c)],
\begin{eqnarray} \label{eq:diffusion_definition}
    D(N) := \frac{1}{2} \left[\textsf{Var}[(x(N)] - \textsf{Var}[(x(N-1)] \right],
\end{eqnarray}
and the local exponent $\alpha = \alpha (N)$ [Fig.~\ref{Fig:7}(b)] defined by a discrete logarithmic step derivative,
\begin{eqnarray} \label{eq:local_exponent}
\alpha(N):= \frac{\ln \left[\ \textsf{Var}[x(N)] \right] - \ln \left[ \textsf{Var}[x(N-1)] \right]}{\ln \left[ N \right] - \ln \left[ N - 1 \right]}.
\end{eqnarray}
Thus, ordinary diffusion corresponds to $\alpha=1$, whereas local subdiffusive and superdiffusive behavior is indicated by $\alpha<1$ and $\alpha>1$, respectively.  Transport at strong driving is dominated by a superdiffusive regime which grows with increasing strength of the driving [Fig.~\ref{Fig:7}(b)] while the velocity keeps dropping [Fig.~\ref{Fig:7}(a)]
% %%%%%%%%%%%%%%%%%%%%%%%%%%%%
% %%%%%%%%%%%%%%%%%%%%%%%%%%%%%

The superdiffusion for stronger forces [Fig.~\ref{Fig:7}(b)] can be rationalized as follows. While the average velocity keeps dropping with increasing force [Fig.~\ref{Fig:7}(a)], there is a disparity between the trajectories of the tracer that have not yet hit an obstacle and follow a free path, with those that are stuck on an obstacle. Initially, this disparity is increased with the number of steps as indicated by a discrete time window of superdiffusion with a growing exponent $\alpha > 1$ in Fig.~\ref{Fig:7}(a). Once we reach discrete time-scales that are much larger than the time of the mean-free path, regular diffusion is recovered but with a significantly larger diffusion coefficient compared to the bare dynamics. This happens as long as the tracer can eventually go around the obstacle. 
For increasing forces, the number of steps it takes to go around an obstacle is increased, which in turn increases this disparity since the trajectory with the free path can cover a larger distance during this time. 

To investigate this effect, we  develop an asymptotic model similar to Ref.~\cite{leitmann2017time}. At large forces $F\gg 1$, the tracer's trajectory can be approximated as performing jumps only in the directions of the force in one dimension until it hits an obstacle. Once this happens, the tracer is stuck and stays there. Therefore the asymptotic model is the following: at each step the tracer has a probability to step along the direction of the force and hit an obstacle with probability $p = n$ or move forward with probability $1-n$. The probability of displacement $x = j$ after $N$ jumps now reads
\begin{align}
   \mathbb{P}[x=j | N] =& q^j p+q^j(1-p) \delta_{j N} \nonumber \\
    =& p+\delta_{j N}[1-(N+1) p]+O\left(p^2\right),\nonumber \\
    &\quad j=0, \ldots, N, \quad q=1-p.
\end{align}
Where we have approximated to first order in the obstacle density $O(p=n)$ by using the approximation $(1-p)^N = 1-N p + O(p^2)$. The first and second moments can now be calculated,
\begin{eqnarray}
\left\langle x(N)\right\rangle&=&\sum_{j=0}^N j \mathbb{P}[x=j|N]\nonumber \\
&=&\frac{1}{2} n N(N+1)+ N[1-(N+1) n]+O\left(n^2\right),\nonumber \\
\end{eqnarray}
\begin{eqnarray}
\left\langle x(N)^2\right\rangle &=& \sum_{j=0}^N j^2 \mathbb{P}(x=j|N)\nonumber \\
&=&\frac{1}{6} n N(N+1)(2 N+1) +N^2[1-(N+1) n] \nonumber \\
&& +O\left(n^2\right).
\end{eqnarray}
Hence the variance is
\begin{align} \label{eq:var_superdiffusion}
    \textsf{Var} [x(N)] =& \frac{1}{2} (1+n) N \left[1-\tanh ^2\left(\frac{F}{4}\right)\right] \nonumber \\
    &+ n\frac{N}{6} + n\frac{N^2}{2} + n\frac{N^3}{3} + O(n^2),
\end{align}
corrected by the empty lattice (the first term) since for $F\gg 1$, $\tanh (F/4) \approx1$, and it drops, but in simulations [Fig.~\ref{Fig:7}] we still plot the behavior for finite values of $F$.
The diffusion coefficient from Eq.~\eqref{eq:diffusion_definition} is now
\begin{align} \label{eq:diffusion_asymptotic}
    D(N) = \frac{1}{4}(1+n)\left[ 1- \tanh^2 \left( \frac{F^2}{4} \right) \right] + \frac{1}{2} n (N+1)^2. 
\end{align}
Equation~\eqref{eq:var_superdiffusion} suggests that the true exponent of superdiffusion is $\alpha=3$ as is corroborated by simulations in Fig.~\ref{Fig:7}. The reason for the decay of $\alpha$ to unity  at large steps $N$, is the fact that eventually for any finite forces $F$ the tracer can go around the obstacle if we wait long enough. Therefore, regular diffusion $\alpha = 1$ is eventually regained for $N \gg 1$ but with a considerably higher diffusion coefficient [Fig.~\ref{Fig:7}(c)] compared to the bare one defined by $D_0=[1 - \tanh ^2 (F/4)] / 4$.

\section{The diffusion coefficient in the regime of large steps } \label{sec:diffusion_asymptotic}
We now determine the diffusion coefficient for large step numbers, $N \gg 1$, defined using the variance by $D_{\infty} = \lim_{N\to \infty} \textsf{Var}[x(N)]/2N$.
We rewrite the MSD in this regime, which was found in Eq.~\eqref{eq:MSD}, as
\begin{eqnarray}
    \langle x(N)^2 \rangle &=& (1+n) (2 D_0 N + v_0^2 N^2) \nonumber \\
    &&+n \left[ 2 v_0 \Delta v_{\infty} N^2 + v_0^2 N^2 + c_1(F) N \right],
\end{eqnarray}
where $D_0 = \left[ 1 - \tanh ^2 (F/4)\right] / 4$ is the diffusion coefficient of the bare dynamics in the direction of $F$.
Expanding now $\langle x(N) \rangle^2$ from Eq.~\eqref{eq:x(N)} to first order in the obstacle density $n$ we find,
\begin{eqnarray}
    \langle x(N) \rangle^2 &=& [v_0 + n (v_0 + \Delta v_{\infty})]^2 N^2 \nonumber \\
    &=& [v_0^2 + 2 n v_0(v_0 + \Delta v_{\infty})] N^2 +O(n^2).
\end{eqnarray}
The diffusion coefficient is now obtained,
\begin{eqnarray} \label{eq:terminal_variance}
    D_\infty &=& \left[ \langle x(N)^2 \rangle - \langle x(N) \rangle^2 \right] / N \nonumber \\
    &=& (1+n) D_0 + n \, c_1(F) /2 + O(n^2).
\end{eqnarray}
We plot the case of $F=0$ in Fig.~\ref{Fig:4} where $c_1(F=0)= -\pi / 2$. The full function $c_1(F)$ is given in the appendix and in Fig.~\ref{Fig:8} we plot $D_\infty$ in the regime of large steps. Higher-order corrections show that there is a complex non-linear dependence on the force, specifically the logarithmic dependence on $F$ via $c_1 \approx -\pi /2 - (0.33 + 0.11 \ln (F)) F^2 + O(F^4)$, as was also mentioned previously in Sec.~\ref{sec:second_moment_limit}.

% %%%%%%%%%%%%%%%%%%%%%%%%%%%%
\begin{figure}[!htb]
\centering
\includegraphics[width=0.45\textwidth]{"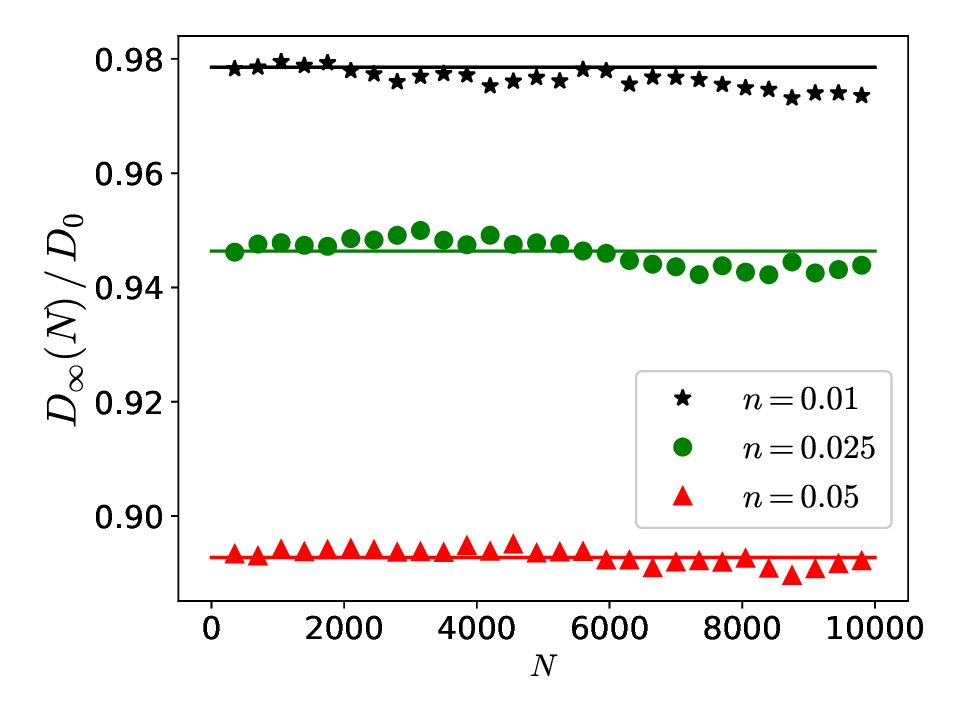"}
\caption{The behavior of the diffusion coefficient $D_\infty$ in the regime of large steps $N$ divided by the diffusion $D_0$ of the bare dynamics (no obstacles) for different obstacle densities $n$. The force is $F=0.1$. Solid lines are the analytical solution for in Eq.~\eqref{eq:terminal_variance} and the symbols are the simulations results. The trend of lowered diffusion (compared to theory) in the $N>5000$ region, is probably due to the effects of starting to reach the boundaries of the system (our lattice is periodic with size $1000 \times 1000$) which were ignored in our theory.
}
\label{Fig:8}
\end{figure}
% %%%%%%%%%%%%%%%%%%%%%%%%%%%%%

\section{Summary and conclusions} \label{section:discussion}
In this work we have considered the driven Lorentz model of a tracer particle hopping on a two-dimensional lattice where a fraction of the sites is inaccessible and acts as hard obstacles.
We derive a method to find the exact solution for the first and second moment of the displacement in $z$-space [Eq.~\eqref{eq:x(z)} and Eq.~\eqref{eq:x^2(Z)}] to first order in the obstacle density $n$ for any value of $z$ (the discrete Laplace transform) and force $F$.
Our result is correct to first order in the obstacle densities where interactions with single obstacles at a time dominate the process. 
Meaning, obstacles are assumed to be far enough from each other so the effects of trapping by a cluster of obstacles on the tracer can be ignored. 
Considering the mean displacement as a function of steps, $\langle x(N) \rangle$, the first-order expansion in the force obeys linear response in terms of the obstacle density [Eq.~\eqref{eq:full_first_moment}], consistent with previous works~\cite{leitmann2013nonlinear, nieuwenhuizen1986diffusion}.
The next correction term to the terminal velocity already includes a logarithm, $v_\infty = D_x F + (n/16) (\pi/4 -1) F^3 \ln (F) + O(F^3)$.
Simulations [Fig.~\ref{Fig:2}-\ref{Fig:8}] show a good match with theory even for low values of the number of steps.
We note that the point where the theory breaks down depends both on the obstacle density and the magnitude of the force as can be seen the simulation results.
For  larger values of $F$, the range of validity of our theory shrinks to smaller and smaller obstacle densities. Reversely, the role of clusters becomes more and more relevant, the larger force is.

By switching to continuous time with an exponential waiting time PDF we are able to recover all previous result from \cite{leitmann2013nonlinear}. 
This is expected since for finite average waiting times $\tau$, the tracer on average performs a jump every $\tau$ (arbitrary units), and the connection between $N$ and $t$ is $N = t / \tau$ (for large values of $N$).
It was predicted that for exponential waiting times the obstacles introduce anti-correlations into the system that break the power-law decay $\sim t^{-1}$ towards the terminal velocity expected from linear response \cite{ernst1971long, nieuwenhuizen1987density}.
In particular the decay has an exponential dependence on the force and acts as $\sim t^{-1} \exp (-F^2 t / 16)$ for $F \to 0$ \cite{leitmann2013nonlinear}.
This is consistent with the domain of discrete time, where each step is distributed according to a delta function, we find $\sim N^{-1 } \exp (-F^2 N / 16)$ with an onset step-time of $N_f \sim 16/F^2$ for small $F$ (see Sec.~\ref{sec:decay_rate}).

Investigating the behavior of the variance at an intermediate number of steps shows non-monotonic behavior in $N$ even at small values of force $F$ [Fig~\ref{Fig:6}]. For larger forces there is a window of superdiffusion at intermediate values of steps $N$ where the variance behaves as $\textsf{Var}[x(N)] \propto N^3$ [Eq.\eqref{eq:var_superdiffusion}] until it drops to regular diffusion with a linear dependence on $N$ [Fig.~\ref{Fig:7}].
In the lattice Lorentz model, the superdiffusion can be traced back to the rapid increase of the variance of the free path lengths as the tracer performs a purely directed motion along the field until it hits an obstacle.
Although increasing the applied force on the tracer can reduce its travel time between different obstacles, it will increase the time it spends trapped by the obstacles.
Such non-monotonic behavior and superdiffusivity has been found in previous works for the continuous case of exponentially distributed times between steps~\cite{illien2018nonequilibrium, leitmann2017time, illien2014velocity, benichou2014microscopic}. 
The superdiffusion regime starts immediately for discrete time [Fig.~\ref{Fig:7}(b)], 
while in the exponential waiting time case there is still a subdiffusive regime at small number of steps of the order of the density $n$~\cite{leitmann2017time}.
Our results show that a superdiffusive regime is a generic feature occurring naturally in the lattice Lorentz gas regardless of the time statistics between steps.

The developed methodology can be used in future research to consider systems where different types of disorder exist, such as non-static obstacles or the quenched temporal disorder that is present in the quenched trap model~\cite{akimoto2020trace,shafir2022case,shafir2024disorder}. 

\begin{acknowledgements}
This work was supported by the  Israel Science Foundation Grant No. 2796/20. AS acknowledges FWF Der Wissenschaftsfonds for funding through the Lise-Meitner Fellowship (Grant DOI 10.55776/M3300). TF gratefully acknowledges support by the Austrian Science Fund (FWF) (Grant DOI 10.55776/M3300).
\end{acknowledgements}

\appendix 
\section{The obstacle-free propagator $\hat{G}_0(z)$} \label{sec:appendix}
\subsection{The matrix elements} \label{sec:the_matrix_elements}
We explicitly evaluate the elements of the matrix $\hat{G}_0(z)$ using a representation that makes use of $I_m(\dots)$, a modified Bessel function of the first kind of integer order $m$.
We use an integral representation of a fraction, $\int_0^{\infty} ds e^{-s a} = 1/a$ for $a>0$, to convert Eq.~\eqref{eq:z-transform_definition} into a  more familiar form
%%%%%%%%%%%%%%%%%%%%%
\begin{eqnarray} \label{eq:G0_definition}
\langle \mathbf{r} | \hat{G}_0(z) | \mathbf{r'} \rangle &=& \int_{-\pi}^{\pi} \frac{dk_x}{2 \pi} e^{-i k_x (x-x')} \nonumber  \\
& & \times \int_{-\pi}^{\pi} \frac{dk_y}{2 \pi}  e^{-i k_y (y-y')} \int_0^{\infty} ds e^{-s[1-z \lambda (\mathbf{k})]} \nonumber . \\
\end{eqnarray}
%%%%%%%%%%%%%%%%%%%%%%%%
Substituting Eq.~\eqref{eq:lambda_biased} into Eq.~\eqref{eq:G0_definition} we obtain 
%%%%%%%%%%%%%%%%%%%%
\begin{eqnarray} 
    &\langle \mathbf{r} | \hat{G}_0(z) | \mathbf{r'} \rangle = \nonumber\\
    &\int_{0}^{\infty} ds e^{-s} \int_{-\pi}^{\pi} \frac{dk_x}{2 \pi} \exp \Big[2sz \Gamma \Big( \cos (k_x) \cosh (F/2) \nonumber \\
    &+ i \sin (k_x) \sinh (F/2) \Big) - i k_x ( x - x' ) \Big] \nonumber \\
    & \times \int_{-\pi}^{\pi} \frac{dk_y}{2 \pi} \exp [2s z \Gamma \cos (k_y) -i k_y ( y - y' ) ] .
\end{eqnarray}
%%%%%%%%%%%%%%%%
We then use the relation (\cite{montroll1973random}, Eq. (64))
%%%%%%%%%%%%%%%%%%%%%%%%%%%%
\begin{eqnarray}
    &&\int_{-\pi}^\pi \frac{\mathrm{d} k}{2 \pi} \exp [-i k m] \exp [\alpha \cos (k)+ i \beta \sin (k)]= \nonumber \\
    &&\left[\frac{\alpha+\beta}{\sqrt{\alpha^2-\beta^2}}\right]^m I_m\left(\sqrt{\alpha^2-\beta^2}\right).
\end{eqnarray}
%%%%%%%%%%%%%%%%%%%%%%%%%%%%%
Notice that the positions $x$ and $y$ were defined to be an integer number (between 1 and $L$) in Sec.~\ref{sec:the_model}. 
Therefore the matrix elements of the free propagator are
%%%%%%%%%%%%%%%%%%%%%%
\begin{eqnarray}
    \langle \mathbf{r} | \hat{G}_0(z) | \mathbf{r'} \rangle &=& e^{F (x - x') / 2} \int_0^{\infty} d s e^{-s} \\
    & & \times   I_{|x-x'|}\left(2 \Gamma s z \right) I_{|y-y'|}\left(2 \Gamma s z\right) , \nonumber
\end{eqnarray}
%%%%%%%%%%%%%%%%%%%%%%
where the exponential outside of the integral accounts for the asymmetry introduced by the bias and the remaining terms are the solution of the symmetrical problem but with the normalization factor set to be $\Gamma(F)$ instead of $\Gamma(F=0)$. 
\subsection{Force-free propagators relations} \label{sec:force_free_prop_relations}
We show that the four terms of the free symmetrical propagates, $g_{00}$, $g_{10}$, $g_{11}$, $g_{20}$, appearing in the matrix elements of $\hat{G}_0(z)$ [see Eq. \ref{eq:G0_matrix_form}],
\begin{equation}
\langle \mathbf{r} | \hat{G}_0(z) | \mathbf{r'} \rangle = e^{F (x - x') / 2} g_{x-x',y-y'},
\end{equation}
are related to each another.
Thus only the knowledge of two is required to know the other two as well.
The symmetrical case $g_{xy}$ with \textit{no} obstacles in Eq.~\eqref{eq:g_formula} from its definition is also represented in the form~\cite{weiss1994aspects, klafter2011first, hughes1995random}
\begin{equation}
    g_{xy} = \sum_{N=0}^{\infty} z^N P_N(\mathbf{r}) ,
\end{equation}
where $\mathbf{r}=(x,y)$ and just for convenience we switch to the notation that $P_N(\mathbf{r}) = \langle \mathbf{r} | p_n \rangle$ is the probability to be at position $\mathbf{r}$ after $N$ steps, given we started at zero. 
The process is translationally invariant and we are free to choose our starting condition to be at the origin, $P_0(\mathbf{r}) = \delta_{\mathbf{r},\mathbf{0}}$ 
(here $\delta$ denotes the Kronecker symbol)
the system obeys a Markovian process 
\begin{equation}
    P_{N+1}(\mathbf{r}) = \sum_{\mathbf{d} \in \mathcal{N}} W(\mathbf{d}) P_N (\mathbf{r} - \mathbf{d}) ,
\end{equation}
multiplying by $z^N$ and summing for $N$ from $0$ to $\infty$
\begin{equation}
    \frac{1}{z} \sum_{N=1}^{\infty} z^N P_N(\mathbf{r}) = \sum_{\mathbf{d} \in \mathcal{N}} W(\mathbf{d}) g_{\mathbf{r-d}} ,
\end{equation}
now using $P_0(\mathbf{r}) = \delta _{\mathbf{r}, \mathbf{0}}$ we obtain
\begin{equation}
    g_{\mathbf{r}} = \delta _{\mathbf{r}, \mathbf{0}} + z \sum_{\mathbf{d} \in \mathcal{N}} W(\mathbf{d}) g_{\mathbf{r-d}}.
\end{equation}
And using $\mathbf{r} = \mathbf{0}$ and $\mathbf{r} = (1,0)$ we find the relations 
\begin{equation}
    \begin{aligned}
        g_{00} &= 1 + 4 \Gamma z g_{10} \\
        g_{10} &= \Gamma z ( g_{00} + g_{20} + 2 g_{11} ).
    \end{aligned}
\end{equation}

\section{Tauberian theorem}
The Tauberian Theorem \cite{klafter2011first, weiss1994aspects, hughes1995random} allows transitioning to $N$-space and find $\langle x(N) \rangle$ for large $N$ from the behavior of its generating function $\langle \tilde{x} (z) \rangle =\sum_{N=0}^{\infty} \langle x(N) \rangle z^N$ when $z\to 1$.
To be more precise, given that near $z \to 1$ the function behaves as
\begin{equation}
    \langle \tilde{x} (z) \rangle \sim \frac{1}{(1-z)^\gamma} Y\left(\frac{1}{1-z}\right),
\end{equation}
where $\gamma$ is some positive number and $Y(u)$ is a slowly varying function of $u$, i.e.,
\begin{equation}
\lim _{u \rightarrow \infty} \frac{Y(C u)}{Y(u)}=1,
\end{equation}
for any positive constant $C$.
If the sequence $\{ \langle x(N) \rangle \}$ is monotonic (at least starting from some value of $N$) then
\begin{equation}
\langle x(N) \rangle \cong \frac{1}{\Gamma(\gamma)} N^{\gamma-1} Y(N),
\end{equation}
where $\Gamma$ is the Gamma-function.
For example the $1/(1-z)^2$ term in Eq. (\ref{eq:x(z)}) is transformed to $N$  using $\gamma =2$ and $Y(u)=1$ and we obtain
%%%%%%%%%%%%%%%%%%%%%%%%%%%%%%%
\begin{eqnarray}
    \langle x(N) \rangle &=& (1+n)\langle x_0(N) \rangle - n \frac{\pi F}{4} N \\ \nonumber
    &=&  \Big(1 - n (\pi -1)\Big)\frac{F}{4} N \\ \nonumber
    &=& D_x F N
\end{eqnarray}
%%%%%%%%%%%%%%%%%%%%%%%%%%%%%%%
to first order in the force $F$ and the regime $N \gg 1$ correct to first order in the obstacle density $n$.

\section{Asymptotic behavior for $z\to 1$} \label{section:asymptotic_limits}
In this section we provide the exact asymptotics ($z \to 1$) of the scattering matrix for any force $F$, used to determine the stationary velocity $v_{\infty}$ and diffusion coefficient $D_\infty$ appearing in the figures.

The first derivative of $t_i(z, \mathbf{k})$ can be computed with Wolfram Mathematica asymptotically for $z \to 1$ with the result
%%%%%%%%%%%%%%%%%%%%%%%%%%%%%%%%%%%
\begin{widetext}
\begin{eqnarray} \label{eq:full_first_moment}
    && \left.\left(-\mathrm{i} L^2 \frac{\partial t_i(z, \mathbf{k})}{\partial k_x} \right) \right|_{\mathbf{k}=0} = \\ \nonumber
    && -\frac{4 (\pi -2 \mathrm{E}) \sinh \left(\frac{F}{2}\right) \left[(\pi -2 \mathrm{E}) \cosh ^4\left(\frac{F}{4}\right)+2 \mathrm{K} \sinh ^2\left(\frac{F}{4}\right)\right]}{\left[8 \mathrm{E} \cosh ^4\left(\frac{F}{4}\right)-\mathrm{K} \left(4 \cosh \left(\frac{F}{2}\right)+\cosh (F)-5\right)\right] \left(-2 \mathrm{E} \cosh ^2\left(\frac{F}{4}\right)+2 \mathrm{K} \sinh ^2\left(\frac{F}{4}\right)+\pi \right)}-2 \tanh \left(\frac{F}{4}\right) +O(1-z) \\ \nonumber
    &&= \Delta v_{\infty} + O(1-z),
\end{eqnarray}
\end{widetext}
%%%%%%%%%%%%%%%%%%%%%%%%%%%%%%%%%%%%
where we abbreviated $\mathrm{K} = \boldsymbol{K}(\sech^4 (F / 4))$ and $\mathrm{E} = \boldsymbol{E}(\sech^4 (F/4))$ and the derivative of $t_i(z,\mathbf{k})$ was evaluated up to terms of order $O(1-z)$.

Similarly, the second derivative of $t_i(z, \mathbf{k})$ can be expanded using computer algebra for $z\to 1$ with the result
%%%%%%%%%%%%%%%%%%%%%%%%%%%%%%%%%%%%%%%
\begin{widetext}
\begin{eqnarray} \label{eq:full_second_moment} 
& & \left.L^2 \frac{\partial^2}{\partial^2 k_x} {t}_i (z, \mathbf{k}) \right|_{\mathbf{k}=0} = \\ \nonumber
& & \frac{\text{sech}^2\left(\frac{F}{4}\right) \left[U_1 + \pi \mathrm{K} U_2 +2 \mathrm{E} U_3 +\pi ^3 \left(8 \cosh \left(\frac{F}{2}\right)+5 \cosh (F)-5\right) \cosh ^4\left(\frac{F}{4}\right)\right]}{2 \pi  \left[8 \cosh ^4\left(\frac{F}{4}\right) \mathrm{E}-\left(4 \cosh \left(\frac{F}{2}\right)+\cosh (F)-5\right) \mathrm{K}\right] \left(2 \sinh ^2\left(\frac{F}{4}\right) \mathrm{K}-2 \cosh ^2\left(\frac{F}{4}\right) \mathrm{E}+\pi \right)}\\ \nonumber
& &  -2 \tanh ^2\left(\frac{F}{4}\right) \frac{z}{1-z} + O(1-z)\\ \nonumber
&& = c_1(F) -2 (v_0)^2 \frac{z}{1-z} +O(1-z).
\end{eqnarray}
\end{widetext}
%%%%%%%%%%%%%%%%%%%%%%%%%%%%%%%%%%%%%%%
where
\begin{widetext}
\begin{eqnarray}
    U_1 &=& 128 \sinh ^2\left(\frac{F}{4}\right) \cosh ^6\left(\frac{F}{4}\right) \mathrm{K} \mathrm{E}^2,\\
    U_2 &=& \pi  \left(7 \cosh \left(\frac{F}{2}\right)+\cosh \left(\frac{3 F}{2}\right)-6 \cosh (F)-2\right)-4 \sinh ^4\left(\frac{F}{4}\right) \left(16 \cosh \left(\frac{F}{2}\right)+3 \cosh (F)-3\right) \mathrm{K} ,\\
    U_3 &=& 6 \pi  \sinh ^4\left(\frac{F}{2}\right) \mathrm{K}-16 \sinh ^4\left(\frac{F}{4}\right) \left(\cosh \left(\frac{F}{2}\right)+3\right) \cosh ^2\left(\frac{F}{4}\right) \mathrm{K}^2+\pi ^2 (1-9 \cosh (F)) \cosh ^4\left(\frac{F}{4}\right).     
\end{eqnarray}
\end{widetext}
For small $0 < F\ll 1 $ this can be further simplified to fourth order in the force to
\begin{equation} \label{eq:second_derivative_t}
\left.L^2 \frac{\partial^2}{\partial^2 k_x} t_i (z, \mathbf{k}) \right|_{\vec{k}=0}=c_1(F) - \left(\frac{F^2}{8}\right) \frac{z}{1-z} + O(1-z).
\end{equation}
where $c_1(F)$ in the regime of small $F$ takes the form,
\begin{widetext}
    \begin{eqnarray}
        c_1 &=& \frac{F^2}{64 \pi (\pi -2)} \left\{ 2 \left[\pi ^2 (\pi -6)+16\right] \log (F)-112 \log (2)+\pi  \{16+\pi  [22+42 \log (2)-\pi  (13+\log (128))]\} \right\} \nonumber \\
    && -\frac{\pi}{2} + O(F^4).
    \end{eqnarray}
\end{widetext}

\section{Simulations} \label{sec:simulations}
In this section we provide details on the method of simulations.
Fluctuations in the bare dynamics (without obstacles) are much stronger than the obstacle induced ones, especially when the number of steps is small. 
% Therefore we need to average over a larger ensemble of trajectories for simulations to converge towards theory when compared to the large $N$ limit.
Therefore to construct the simulations, it is very useful to adapt the approach from ref.~\cite{frenkel1987velocity} which was also successfully adapted in ref.~\cite{rusch2024noise, chambers1976method}.
Let us denote the difference in position of a test particle for the same sequence of trial moves with and without obstacles (excluded sites) by $\delta x$. Clearly, the average magnitude of $\delta x$ is proportional to the density of obstacles. Let us write the total position of a particle in the presence of hard obstacles as 
\begin{equation}
    x(N) = x_0(N) + \delta x.
\end{equation}
$\langle \delta x \rangle$ can be computed directly from simulations and converges much faster than $\langle x(N) \rangle$ while $x_0(N)$ is the position of the bare-dynamics (no obstacles) which is known. 
To further speed up simulations in the small-$N$ regime, we use a second trick. We simulate obstacle positions only in the effective region of the trajectory. Meaning, obstacles only appear in the area the test particle can reach up to step number $N$. We then normalize the result to account for the shifted dynamics by using the fact that for obstacle outside the effective region, the mean position would just be that of the bare dynamics $x_0(N)$. 
Same methodology were used to determine $x(N)^2$.

%\bibliographystyle{naturemag}
% \bibliographystyle{apsrev4-1-title_noeprint}

% \nocite{*} 
\bibliography{./main.bib}

\begin{thebibliography}{59}%
\makeatletter
\providecommand \@ifxundefined [1]{%
 \@ifx{#1\undefined}
}%
\providecommand \@ifnum [1]{%
 \ifnum #1\expandafter \@firstoftwo
 \else \expandafter \@secondoftwo
 \fi
}%
\providecommand \@ifx [1]{%
 \ifx #1\expandafter \@firstoftwo
 \else \expandafter \@secondoftwo
 \fi
}%
\providecommand \natexlab [1]{#1}%
\providecommand \enquote  [1]{``#1''}%
\providecommand \bibnamefont  [1]{#1}%
\providecommand \bibfnamefont [1]{#1}%
\providecommand \citenamefont [1]{#1}%
\providecommand \href@noop [0]{\@secondoftwo}%
\providecommand \href [0]{\begingroup \@sanitize@url \@href}%
\providecommand \@href[1]{\@@startlink{#1}\@@href}%
\providecommand \@@href[1]{\endgroup#1\@@endlink}%
\providecommand \@sanitize@url [0]{\catcode `\\12\catcode `\$12\catcode `\&12\catcode `\#12\catcode `\^12\catcode `\_12\catcode `\%12\relax}%
\providecommand \@@startlink[1]{}%
\providecommand \@@endlink[0]{}%
\providecommand \url  [0]{\begingroup\@sanitize@url \@url }%
\providecommand \@url [1]{\endgroup\@href {#1}{\urlprefix }}%
\providecommand \urlprefix  [0]{URL }%
\providecommand \Eprint [0]{\href }%
\providecommand \doibase [0]{https://doi.org/}%
\providecommand \selectlanguage [0]{\@gobble}%
\providecommand \bibinfo  [0]{\@secondoftwo}%
\providecommand \bibfield  [0]{\@secondoftwo}%
\providecommand \translation [1]{[#1]}%
\providecommand \BibitemOpen [0]{}%
\providecommand \bibitemStop [0]{}%
\providecommand \bibitemNoStop [0]{.\EOS\space}%
\providecommand \EOS [0]{\spacefactor3000\relax}%
\providecommand \BibitemShut  [1]{\csname bibitem#1\endcsname}%
\let\auto@bib@innerbib\@empty
%</preamble>
\bibitem [{\citenamefont {Scher}\ and\ \citenamefont {Montroll}(1975)}]{scher1975anomalous}%
  \BibitemOpen
  \bibfield  {author} {\bibinfo {author} {\bibfnamefont {H.}~\bibnamefont {Scher}}\ and\ \bibinfo {author} {\bibfnamefont {E.~W.}\ \bibnamefont {Montroll}},\ }\bibfield  {title} {\bibinfo {title} {Anomalous transit-time dispersion in amorphous solids},\ }\href {https://doi.org/10.1103/PhysRevB.12.2455} {\bibfield  {journal} {\bibinfo  {journal} {Physical Review B}\ }\textbf {\bibinfo {volume} {12}},\ \bibinfo {pages} {2455} (\bibinfo {year} {1975})}\BibitemShut {NoStop}%
\bibitem [{\citenamefont {Shlesinger}(1974)}]{shlesinger1974asymptotic}%
  \BibitemOpen
  \bibfield  {author} {\bibinfo {author} {\bibfnamefont {M.~F.}\ \bibnamefont {Shlesinger}},\ }\bibfield  {title} {\bibinfo {title} {Asymptotic solutions of continuous-time random walks},\ }\href {https://doi.org/10.1007/BF01008803} {\bibfield  {journal} {\bibinfo  {journal} {Journal of Statistical Physics}\ }\textbf {\bibinfo {volume} {10}},\ \bibinfo {pages} {421} (\bibinfo {year} {1974})}\BibitemShut {NoStop}%
\bibitem [{\citenamefont {Klafter}\ and\ \citenamefont {Silbey}(1980)}]{klafter1980derivation}%
  \BibitemOpen
  \bibfield  {author} {\bibinfo {author} {\bibfnamefont {J.}~\bibnamefont {Klafter}}\ and\ \bibinfo {author} {\bibfnamefont {R.}~\bibnamefont {Silbey}},\ }\bibfield  {title} {\bibinfo {title} {Derivation of the continuous-time random-walk equation},\ }\href {https://doi.org/10.1103/PhysRevLett.44.55} {\bibfield  {journal} {\bibinfo  {journal} {Physical Review Letters}\ }\textbf {\bibinfo {volume} {44}},\ \bibinfo {pages} {55} (\bibinfo {year} {1980})}\BibitemShut {NoStop}%
\bibitem [{\citenamefont {Scher}(1991)}]{scher1991time}%
  \BibitemOpen
  \bibfield  {author} {\bibinfo {author} {\bibfnamefont {H.}~\bibnamefont {Scher}},\ }\bibfield  {title} {\bibinfo {title} {Time-scale invariance in transport and relaxation},\ }\href {http://www.scepticalphysiologist.com/projectsskills/literature/1991_PhysicsToday_44_26_Scher_REV.pdf} {\bibfield  {journal} {\bibinfo  {journal} {Physics Today}\ ,\ \bibinfo {pages} {26}} (\bibinfo {year} {1991})}\BibitemShut {NoStop}%
\bibitem [{\citenamefont {Metzler}\ \emph {et~al.}(2022)\citenamefont {Metzler}, \citenamefont {Rajyaguru},\ and\ \citenamefont {Berkowitz}}]{metzler2022modelling}%
  \BibitemOpen
  \bibfield  {author} {\bibinfo {author} {\bibfnamefont {R.}~\bibnamefont {Metzler}}, \bibinfo {author} {\bibfnamefont {A.}~\bibnamefont {Rajyaguru}},\ and\ \bibinfo {author} {\bibfnamefont {B.}~\bibnamefont {Berkowitz}},\ }\bibfield  {title} {\bibinfo {title} {Modelling anomalous diffusion in semi-infinite disordered systems and porous media},\ }\href {https://doi.org/10.1088/1367-2630/aca70c} {\bibfield  {journal} {\bibinfo  {journal} {New journal of physics}\ }\textbf {\bibinfo {volume} {24}},\ \bibinfo {pages} {123004} (\bibinfo {year} {2022})}\BibitemShut {NoStop}%
\bibitem [{\citenamefont {Waigh}\ and\ \citenamefont {Korabel}(2023)}]{waigh2023heterogeneous}%
  \BibitemOpen
  \bibfield  {author} {\bibinfo {author} {\bibfnamefont {T.~A.}\ \bibnamefont {Waigh}}\ and\ \bibinfo {author} {\bibfnamefont {N.}~\bibnamefont {Korabel}},\ }\bibfield  {title} {\bibinfo {title} {Heterogeneous anomalous transport in cellular and molecular biology},\ }\href {https://doi.org/10.1088/1361-6633/ad058f} {\bibfield  {journal} {\bibinfo  {journal} {Reports on Progress in Physics}\ } (\bibinfo {year} {2023})}\BibitemShut {NoStop}%
\bibitem [{\citenamefont {Nissan}\ and\ \citenamefont {Berkowitz}(2018)}]{nissan2018inertial}%
  \BibitemOpen
  \bibfield  {author} {\bibinfo {author} {\bibfnamefont {A.}~\bibnamefont {Nissan}}\ and\ \bibinfo {author} {\bibfnamefont {B.}~\bibnamefont {Berkowitz}},\ }\bibfield  {title} {\bibinfo {title} {Inertial effects on flow and transport in heterogeneous porous media},\ }\href {https://doi.org/10.1103/PhysRevLett.120.054504} {\bibfield  {journal} {\bibinfo  {journal} {Physical Review Letters}\ }\textbf {\bibinfo {volume} {120}},\ \bibinfo {pages} {054504} (\bibinfo {year} {2018})}\BibitemShut {NoStop}%
\bibitem [{\citenamefont {Metzler}\ \emph {et~al.}(2014)\citenamefont {Metzler}, \citenamefont {Jeon}, \citenamefont {Cherstvy},\ and\ \citenamefont {Barkai}}]{metzler2014anomalous}%
  \BibitemOpen
  \bibfield  {author} {\bibinfo {author} {\bibfnamefont {R.}~\bibnamefont {Metzler}}, \bibinfo {author} {\bibfnamefont {J.-H.}\ \bibnamefont {Jeon}}, \bibinfo {author} {\bibfnamefont {A.~G.}\ \bibnamefont {Cherstvy}},\ and\ \bibinfo {author} {\bibfnamefont {E.}~\bibnamefont {Barkai}},\ }\bibfield  {title} {\bibinfo {title} {Anomalous diffusion models and their properties: non-stationarity, non-ergodicity, and ageing at the centenary of single particle tracking},\ }\href {https://doi.org/10.1039/C4CP03465A} {\bibfield  {journal} {\bibinfo  {journal} {Physical Chemistry Chemical Physics}\ }\textbf {\bibinfo {volume} {16}},\ \bibinfo {pages} {24128} (\bibinfo {year} {2014})}\BibitemShut {NoStop}%
\bibitem [{\citenamefont {H{\"o}fling}\ and\ \citenamefont {Franosch}(2013)}]{hofling2013anomalous}%
  \BibitemOpen
  \bibfield  {author} {\bibinfo {author} {\bibfnamefont {F.}~\bibnamefont {H{\"o}fling}}\ and\ \bibinfo {author} {\bibfnamefont {T.}~\bibnamefont {Franosch}},\ }\bibfield  {title} {\bibinfo {title} {Anomalous transport in the crowded world of biological cells},\ }\href {https://doi.org/10.1088/0034-4885/76/4/046602} {\bibfield  {journal} {\bibinfo  {journal} {Reports on Progress in Physics}\ }\textbf {\bibinfo {volume} {76}},\ \bibinfo {pages} {046602} (\bibinfo {year} {2013})}\BibitemShut {NoStop}%
\bibitem [{\citenamefont {Weigel}\ \emph {et~al.}(2011)\citenamefont {Weigel}, \citenamefont {Simon}, \citenamefont {Tamkun},\ and\ \citenamefont {Krapf}}]{weigel2011ergodic}%
  \BibitemOpen
  \bibfield  {author} {\bibinfo {author} {\bibfnamefont {A.~V.}\ \bibnamefont {Weigel}}, \bibinfo {author} {\bibfnamefont {B.}~\bibnamefont {Simon}}, \bibinfo {author} {\bibfnamefont {M.~M.}\ \bibnamefont {Tamkun}},\ and\ \bibinfo {author} {\bibfnamefont {D.}~\bibnamefont {Krapf}},\ }\bibfield  {title} {\bibinfo {title} {Ergodic and nonergodic processes coexist in the plasma membrane as observed by single-molecule tracking},\ }\href {https://doi.org/10.1073/pnas.1016325108} {\bibfield  {journal} {\bibinfo  {journal} {Proceedings of the National Academy of Sciences}\ }\textbf {\bibinfo {volume} {108}},\ \bibinfo {pages} {6438} (\bibinfo {year} {2011})}\BibitemShut {NoStop}%
\bibitem [{\citenamefont {Stefani}\ \emph {et~al.}(2009)\citenamefont {Stefani}, \citenamefont {Hoogenboom},\ and\ \citenamefont {Barkai}}]{stefani2009beyond}%
  \BibitemOpen
  \bibfield  {author} {\bibinfo {author} {\bibfnamefont {F.~D.}\ \bibnamefont {Stefani}}, \bibinfo {author} {\bibfnamefont {J.~P.}\ \bibnamefont {Hoogenboom}},\ and\ \bibinfo {author} {\bibfnamefont {E.}~\bibnamefont {Barkai}},\ }\bibfield  {title} {\bibinfo {title} {Beyond quantum jumps: blinking nanoscale light emitters},\ }\href {https://doi.org/10.1063/1.3086100} {\bibfield  {journal} {\bibinfo  {journal} {Physics Today}\ }\textbf {\bibinfo {volume} {62}},\ \bibinfo {pages} {34} (\bibinfo {year} {2009})}\BibitemShut {NoStop}%
\bibitem [{\citenamefont {Barkai}\ \emph {et~al.}(2012)\citenamefont {Barkai}, \citenamefont {Garini},\ and\ \citenamefont {Metzler}}]{Barkai2012single}%
  \BibitemOpen
  \bibfield  {author} {\bibinfo {author} {\bibfnamefont {E.}~\bibnamefont {Barkai}}, \bibinfo {author} {\bibfnamefont {Y.}~\bibnamefont {Garini}},\ and\ \bibinfo {author} {\bibfnamefont {R.}~\bibnamefont {Metzler}},\ }\bibfield  {title} {\bibinfo {title} {of single molecules in living cells},\ }\href {https://doi.org/10.1063/PT.3.1677} {\bibfield  {journal} {\bibinfo  {journal} {Physics Today}\ }\textbf {\bibinfo {volume} {65}},\ \bibinfo {pages} {29} (\bibinfo {year} {2012})}\BibitemShut {NoStop}%
\bibitem [{\citenamefont {Jeon}\ \emph {et~al.}(2011)\citenamefont {Jeon}, \citenamefont {Tejedor}, \citenamefont {Burov}, \citenamefont {Barkai}, \citenamefont {Selhuber-Unkel}, \citenamefont {Berg-S{\o}rensen}, \citenamefont {Oddershede},\ and\ \citenamefont {Metzler}}]{jeon2011vivo}%
  \BibitemOpen
  \bibfield  {author} {\bibinfo {author} {\bibfnamefont {J.-H.}\ \bibnamefont {Jeon}}, \bibinfo {author} {\bibfnamefont {V.}~\bibnamefont {Tejedor}}, \bibinfo {author} {\bibfnamefont {S.}~\bibnamefont {Burov}}, \bibinfo {author} {\bibfnamefont {E.}~\bibnamefont {Barkai}}, \bibinfo {author} {\bibfnamefont {C.}~\bibnamefont {Selhuber-Unkel}}, \bibinfo {author} {\bibfnamefont {K.}~\bibnamefont {Berg-S{\o}rensen}}, \bibinfo {author} {\bibfnamefont {L.}~\bibnamefont {Oddershede}},\ and\ \bibinfo {author} {\bibfnamefont {R.}~\bibnamefont {Metzler}},\ }\bibfield  {title} {\bibinfo {title} {In vivo anomalous diffusion and weak ergodicity breaking of lipid granules},\ }\href {https://doi.org/10.1103/PhysRevLett.106.048103} {\bibfield  {journal} {\bibinfo  {journal} {Physical Review Letters}\ }\textbf {\bibinfo {volume} {106}},\ \bibinfo {pages} {048103} (\bibinfo {year} {2011})}\BibitemShut {NoStop}%
\bibitem [{\citenamefont {Tabei}\ \emph {et~al.}(2013)\citenamefont {Tabei}, \citenamefont {Burov}, \citenamefont {Kim}, \citenamefont {Kuznetsov}, \citenamefont {Huynh}, \citenamefont {Jureller}, \citenamefont {Philipson}, \citenamefont {Dinner},\ and\ \citenamefont {Scherer}}]{tabei2013intracellular}%
  \BibitemOpen
  \bibfield  {author} {\bibinfo {author} {\bibfnamefont {S.~A.}\ \bibnamefont {Tabei}}, \bibinfo {author} {\bibfnamefont {S.}~\bibnamefont {Burov}}, \bibinfo {author} {\bibfnamefont {H.~Y.}\ \bibnamefont {Kim}}, \bibinfo {author} {\bibfnamefont {A.}~\bibnamefont {Kuznetsov}}, \bibinfo {author} {\bibfnamefont {T.}~\bibnamefont {Huynh}}, \bibinfo {author} {\bibfnamefont {J.}~\bibnamefont {Jureller}}, \bibinfo {author} {\bibfnamefont {L.~H.}\ \bibnamefont {Philipson}}, \bibinfo {author} {\bibfnamefont {A.~R.}\ \bibnamefont {Dinner}},\ and\ \bibinfo {author} {\bibfnamefont {N.~F.}\ \bibnamefont {Scherer}},\ }\bibfield  {title} {\bibinfo {title} {Intracellular transport of insulin granules is a subordinated random walk},\ }\href {https://doi.org/10.1073/pnas.1221962110} {\bibfield  {journal} {\bibinfo  {journal} {Proceedings of the National Academy of Sciences}\ }\textbf {\bibinfo {volume} {110}},\ \bibinfo {pages} {4911} (\bibinfo {year} {2013})}\BibitemShut {NoStop}%
\bibitem [{\citenamefont {He}\ \emph {et~al.}(2008)\citenamefont {He}, \citenamefont {Burov}, \citenamefont {Metzler},\ and\ \citenamefont {Barkai}}]{he2008random}%
  \BibitemOpen
  \bibfield  {author} {\bibinfo {author} {\bibfnamefont {Y.}~\bibnamefont {He}}, \bibinfo {author} {\bibfnamefont {S.}~\bibnamefont {Burov}}, \bibinfo {author} {\bibfnamefont {R.}~\bibnamefont {Metzler}},\ and\ \bibinfo {author} {\bibfnamefont {E.}~\bibnamefont {Barkai}},\ }\bibfield  {title} {\bibinfo {title} {Random time-scale invariant diffusion and transport coefficients},\ }\href {https://doi.org/10.1103/PhysRevLett.101.058101} {\bibfield  {journal} {\bibinfo  {journal} {Physical Review Letters}\ }\textbf {\bibinfo {volume} {101}},\ \bibinfo {pages} {058101} (\bibinfo {year} {2008})}\BibitemShut {NoStop}%
\bibitem [{\citenamefont {Sokolov}(2008)}]{sokolov2008statistics}%
  \BibitemOpen
  \bibfield  {author} {\bibinfo {author} {\bibfnamefont {I.~M.}\ \bibnamefont {Sokolov}},\ }\bibfield  {title} {\bibinfo {title} {Statistics and the single molecule},\ }\href {https://doi.org/10.1103/Physics.1.8} {\bibfield  {journal} {\bibinfo  {journal} {Physics}\ }\textbf {\bibinfo {volume} {1}},\ \bibinfo {pages} {8} (\bibinfo {year} {2008})}\BibitemShut {NoStop}%
\bibitem [{\citenamefont {Shafir}\ and\ \citenamefont {Burov}(2022)}]{shafir2022case}%
  \BibitemOpen
  \bibfield  {author} {\bibinfo {author} {\bibfnamefont {D.}~\bibnamefont {Shafir}}\ and\ \bibinfo {author} {\bibfnamefont {S.}~\bibnamefont {Burov}},\ }\bibfield  {title} {\bibinfo {title} {The case of the biased quenched trap model in two dimensions with diverging mean dwell times},\ }\href {https://doi.org/10.1088/1742-5468/ac52af} {\bibfield  {journal} {\bibinfo  {journal} {Journal of Statistical Mechanics: Theory and Experiment}\ }\textbf {\bibinfo {volume} {2022}},\ \bibinfo {pages} {033301} (\bibinfo {year} {2022})}\BibitemShut {NoStop}%
\bibitem [{\citenamefont {Bouchaud}(1992)}]{bouchaud1992weak}%
  \BibitemOpen
  \bibfield  {author} {\bibinfo {author} {\bibfnamefont {J.-P.}\ \bibnamefont {Bouchaud}},\ }\bibfield  {title} {\bibinfo {title} {Weak ergodicity breaking and aging in disordered systems},\ }\href {https://doi.org/10.1051/jp1:1992238} {\bibfield  {journal} {\bibinfo  {journal} {Journal de Physique I}\ }\textbf {\bibinfo {volume} {2}},\ \bibinfo {pages} {1705} (\bibinfo {year} {1992})}\BibitemShut {NoStop}%
\bibitem [{\citenamefont {Monthus}\ and\ \citenamefont {Bouchaud}(1996)}]{monthus1996models}%
  \BibitemOpen
  \bibfield  {author} {\bibinfo {author} {\bibfnamefont {C.}~\bibnamefont {Monthus}}\ and\ \bibinfo {author} {\bibfnamefont {J.-P.}\ \bibnamefont {Bouchaud}},\ }\bibfield  {title} {\bibinfo {title} {Models of traps and glass phenomenology},\ }\href {https://doi.org/10.1088/0305-4470/29/14/012} {\bibfield  {journal} {\bibinfo  {journal} {Journal of Physics A: Mathematical and General}\ }\textbf {\bibinfo {volume} {29}},\ \bibinfo {pages} {3847} (\bibinfo {year} {1996})}\BibitemShut {NoStop}%
\bibitem [{\citenamefont {Rinn}\ \emph {et~al.}(2000)\citenamefont {Rinn}, \citenamefont {Maass},\ and\ \citenamefont {Bouchaud}}]{rinn2000multiple}%
  \BibitemOpen
  \bibfield  {author} {\bibinfo {author} {\bibfnamefont {B.}~\bibnamefont {Rinn}}, \bibinfo {author} {\bibfnamefont {P.}~\bibnamefont {Maass}},\ and\ \bibinfo {author} {\bibfnamefont {J.-P.}\ \bibnamefont {Bouchaud}},\ }\bibfield  {title} {\bibinfo {title} {Multiple scaling regimes in simple aging models},\ }\href {https://doi.org/10.1103/PhysRevLett.84.5403} {\bibfield  {journal} {\bibinfo  {journal} {Physical Review Letters}\ }\textbf {\bibinfo {volume} {84}},\ \bibinfo {pages} {5403} (\bibinfo {year} {2000})}\BibitemShut {NoStop}%
\bibitem [{\citenamefont {Rinn}\ \emph {et~al.}(2001)\citenamefont {Rinn}, \citenamefont {Maass},\ and\ \citenamefont {Bouchaud}}]{rinn2001hopping}%
  \BibitemOpen
  \bibfield  {author} {\bibinfo {author} {\bibfnamefont {B.}~\bibnamefont {Rinn}}, \bibinfo {author} {\bibfnamefont {P.}~\bibnamefont {Maass}},\ and\ \bibinfo {author} {\bibfnamefont {J.-P.}\ \bibnamefont {Bouchaud}},\ }\bibfield  {title} {\bibinfo {title} {Hopping in the glass configuration space: subaging and generalized scaling laws},\ }\href {https://doi.org/10.1103/PhysRevB.64.104417} {\bibfield  {journal} {\bibinfo  {journal} {Physical Review B}\ }\textbf {\bibinfo {volume} {64}},\ \bibinfo {pages} {104417} (\bibinfo {year} {2001})}\BibitemShut {NoStop}%
\bibitem [{\citenamefont {Berthier}\ and\ \citenamefont {Biroli}(2011)}]{berthier2011theoretical}%
  \BibitemOpen
  \bibfield  {author} {\bibinfo {author} {\bibfnamefont {L.}~\bibnamefont {Berthier}}\ and\ \bibinfo {author} {\bibfnamefont {G.}~\bibnamefont {Biroli}},\ }\bibfield  {title} {\bibinfo {title} {Theoretical perspective on the glass transition and amorphous materials},\ }\href {https://doi.org/10.1103/RevModPhys.83.587} {\bibfield  {journal} {\bibinfo  {journal} {Reviews of Modern Physics}\ }\textbf {\bibinfo {volume} {83}},\ \bibinfo {pages} {587} (\bibinfo {year} {2011})}\BibitemShut {NoStop}%
\bibitem [{\citenamefont {Lorentz}(1905)}]{lorentz1905mouvement}%
  \BibitemOpen
  \bibfield  {author} {\bibinfo {author} {\bibfnamefont {H.~A.}\ \bibnamefont {Lorentz}},\ }\href@noop {} {\emph {\bibinfo {title} {Le mouvement des {\'e}lectrons dans les m{\'e}taux}}}\ (\bibinfo {year} {1905})\BibitemShut {NoStop}%
\bibitem [{\citenamefont {Leitmann}\ \emph {et~al.}(2018)\citenamefont {Leitmann}, \citenamefont {Schwab},\ and\ \citenamefont {Franosch}}]{leitmann2018time}%
  \BibitemOpen
  \bibfield  {author} {\bibinfo {author} {\bibfnamefont {S.}~\bibnamefont {Leitmann}}, \bibinfo {author} {\bibfnamefont {T.}~\bibnamefont {Schwab}},\ and\ \bibinfo {author} {\bibfnamefont {T.}~\bibnamefont {Franosch}},\ }\bibfield  {title} {\bibinfo {title} {Time-dependent perpendicular fluctuations in the driven lattice lorentz gas},\ }\href {https://doi.org/10.1103/PhysRevE.97.022101} {\bibfield  {journal} {\bibinfo  {journal} {Physical Review E}\ }\textbf {\bibinfo {volume} {97}},\ \bibinfo {pages} {022101} (\bibinfo {year} {2018})}\BibitemShut {NoStop}%
\bibitem [{\citenamefont {Leitmann}\ and\ \citenamefont {Franosch}(2017)}]{leitmann2017time}%
  \BibitemOpen
  \bibfield  {author} {\bibinfo {author} {\bibfnamefont {S.}~\bibnamefont {Leitmann}}\ and\ \bibinfo {author} {\bibfnamefont {T.}~\bibnamefont {Franosch}},\ }\bibfield  {title} {\bibinfo {title} {Time-dependent fluctuations and superdiffusivity in the driven lattice lorentz gas},\ }\href {https://doi.org/10.1103/PhysRevLett.118.018001} {\bibfield  {journal} {\bibinfo  {journal} {Physical Review Letters}\ }\textbf {\bibinfo {volume} {118}},\ \bibinfo {pages} {018001} (\bibinfo {year} {2017})}\BibitemShut {NoStop}%
\bibitem [{\citenamefont {B{\'e}nichou}\ \emph {et~al.}(2016)\citenamefont {B{\'e}nichou}, \citenamefont {Illien}, \citenamefont {Oshanin}, \citenamefont {Sarracino},\ and\ \citenamefont {Voituriez}}]{benichou2016nonlinear}%
  \BibitemOpen
  \bibfield  {author} {\bibinfo {author} {\bibfnamefont {O.}~\bibnamefont {B{\'e}nichou}}, \bibinfo {author} {\bibfnamefont {P.}~\bibnamefont {Illien}}, \bibinfo {author} {\bibfnamefont {G.}~\bibnamefont {Oshanin}}, \bibinfo {author} {\bibfnamefont {A.}~\bibnamefont {Sarracino}},\ and\ \bibinfo {author} {\bibfnamefont {R.}~\bibnamefont {Voituriez}},\ }\bibfield  {title} {\bibinfo {title} {Nonlinear response and emerging nonequilibrium microstructures for biased diffusion in confined crowded environments},\ }\href {https://doi.org/10.1103/PhysRevE.93.032128} {\bibfield  {journal} {\bibinfo  {journal} {Physical Review E}\ }\textbf {\bibinfo {volume} {93}},\ \bibinfo {pages} {032128} (\bibinfo {year} {2016})}\BibitemShut {NoStop}%
\bibitem [{\citenamefont {B{\'e}nichou}\ \emph {et~al.}(2014)\citenamefont {B{\'e}nichou}, \citenamefont {Illien}, \citenamefont {Oshanin}, \citenamefont {Sarracino},\ and\ \citenamefont {Voituriez}}]{benichou2014microscopic}%
  \BibitemOpen
  \bibfield  {author} {\bibinfo {author} {\bibfnamefont {O.}~\bibnamefont {B{\'e}nichou}}, \bibinfo {author} {\bibfnamefont {P.}~\bibnamefont {Illien}}, \bibinfo {author} {\bibfnamefont {G.}~\bibnamefont {Oshanin}}, \bibinfo {author} {\bibfnamefont {A.}~\bibnamefont {Sarracino}},\ and\ \bibinfo {author} {\bibfnamefont {R.}~\bibnamefont {Voituriez}},\ }\bibfield  {title} {\bibinfo {title} {Microscopic theory for negative differential mobility in crowded environments},\ }\href {https://doi.org/10.1103/PhysRevLett.113.268002} {\bibfield  {journal} {\bibinfo  {journal} {Physical Review Letters}\ }\textbf {\bibinfo {volume} {113}},\ \bibinfo {pages} {268002} (\bibinfo {year} {2014})}\BibitemShut {NoStop}%
\bibitem [{\citenamefont {Illien}\ \emph {et~al.}(2014)\citenamefont {Illien}, \citenamefont {B{\'e}nichou}, \citenamefont {Oshanin},\ and\ \citenamefont {Voituriez}}]{illien2014velocity}%
  \BibitemOpen
  \bibfield  {author} {\bibinfo {author} {\bibfnamefont {P.}~\bibnamefont {Illien}}, \bibinfo {author} {\bibfnamefont {O.}~\bibnamefont {B{\'e}nichou}}, \bibinfo {author} {\bibfnamefont {G.}~\bibnamefont {Oshanin}},\ and\ \bibinfo {author} {\bibfnamefont {R.}~\bibnamefont {Voituriez}},\ }\bibfield  {title} {\bibinfo {title} {Velocity anomaly of a driven tracer in a confined crowded environment},\ }\href {https://doi.org/10.1103/PhysRevLett.113.030603} {\bibfield  {journal} {\bibinfo  {journal} {Physical Review Letters}\ }\textbf {\bibinfo {volume} {113}},\ \bibinfo {pages} {030603} (\bibinfo {year} {2014})}\BibitemShut {NoStop}%
\bibitem [{\citenamefont {Jack}\ \emph {et~al.}(2008)\citenamefont {Jack}, \citenamefont {Kelsey}, \citenamefont {Garrahan},\ and\ \citenamefont {Chandler}}]{jack2008negative}%
  \BibitemOpen
  \bibfield  {author} {\bibinfo {author} {\bibfnamefont {R.~L.}\ \bibnamefont {Jack}}, \bibinfo {author} {\bibfnamefont {D.}~\bibnamefont {Kelsey}}, \bibinfo {author} {\bibfnamefont {J.~P.}\ \bibnamefont {Garrahan}},\ and\ \bibinfo {author} {\bibfnamefont {D.}~\bibnamefont {Chandler}},\ }\bibfield  {title} {\bibinfo {title} {Negative differential mobility of weakly driven particles in models of glass formers},\ }\href {https://doi.org/10.1103/PhysRevE.78.011506} {\bibfield  {journal} {\bibinfo  {journal} {Physical Review E}\ }\textbf {\bibinfo {volume} {78}},\ \bibinfo {pages} {011506} (\bibinfo {year} {2008})}\BibitemShut {NoStop}%
\bibitem [{\citenamefont {Squarcini}\ \emph {et~al.}(pear{\natexlab{a}})\citenamefont {Squarcini}, \citenamefont {Tinti}, \citenamefont {Illien}, \citenamefont {B{\'e}nichou},\ and\ \citenamefont {Franosch}}]{Squarcini_2024_1}%
  \BibitemOpen
  \bibfield  {author} {\bibinfo {author} {\bibfnamefont {A.}~\bibnamefont {Squarcini}}, \bibinfo {author} {\bibfnamefont {A.}~\bibnamefont {Tinti}}, \bibinfo {author} {\bibfnamefont {P.}~\bibnamefont {Illien}}, \bibinfo {author} {\bibfnamefont {O.}~\bibnamefont {B{\'e}nichou}},\ and\ \bibinfo {author} {\bibfnamefont {T.}~\bibnamefont {Franosch}},\ }\bibfield  {title} {\bibinfo {title} {Dimensional crossover via confinement in the lattice lorentz gas},\ }\href@noop {} {\bibfield  {journal} {\bibinfo  {journal} {to be published}\ } (\bibinfo {year} {to appear}{\natexlab{a}})}\BibitemShut {NoStop}%
\bibitem [{\citenamefont {Squarcini}\ \emph {et~al.}(pear{\natexlab{b}})\citenamefont {Squarcini}, \citenamefont {Tinti}, \citenamefont {Illien}, \citenamefont {B{\'e}nichou},\ and\ \citenamefont {Franosch}}]{Squarcini_2024_2}%
  \BibitemOpen
  \bibfield  {author} {\bibinfo {author} {\bibfnamefont {A.}~\bibnamefont {Squarcini}}, \bibinfo {author} {\bibfnamefont {A.}~\bibnamefont {Tinti}}, \bibinfo {author} {\bibfnamefont {P.}~\bibnamefont {Illien}}, \bibinfo {author} {\bibfnamefont {O.}~\bibnamefont {B{\'e}nichou}},\ and\ \bibinfo {author} {\bibfnamefont {T.}~\bibnamefont {Franosch}},\ }\bibfield  {title} {\bibinfo {title} {Time-dependent dynamics in the confined lattice lorentz gas},\ }\href@noop {} {\bibfield  {journal} {\bibinfo  {journal} {to be published}\ } (\bibinfo {year} {to appear}{\natexlab{b}})}\BibitemShut {NoStop}%
\bibitem [{\citenamefont {Leitmann}\ and\ \citenamefont {Franosch}(2013)}]{leitmann2013nonlinear}%
  \BibitemOpen
  \bibfield  {author} {\bibinfo {author} {\bibfnamefont {S.}~\bibnamefont {Leitmann}}\ and\ \bibinfo {author} {\bibfnamefont {T.}~\bibnamefont {Franosch}},\ }\bibfield  {title} {\bibinfo {title} {Nonlinear response in the driven lattice lorentz gas},\ }\href {https://doi.org/10.1103/PhysRevLett.111.190603} {\bibfield  {journal} {\bibinfo  {journal} {Physical Review Letters}\ }\textbf {\bibinfo {volume} {111}},\ \bibinfo {pages} {190603} (\bibinfo {year} {2013})}\BibitemShut {NoStop}%
\bibitem [{\citenamefont {Basu}\ and\ \citenamefont {Maes}(2014)}]{basu2014mobility}%
  \BibitemOpen
  \bibfield  {author} {\bibinfo {author} {\bibfnamefont {U.}~\bibnamefont {Basu}}\ and\ \bibinfo {author} {\bibfnamefont {C.}~\bibnamefont {Maes}},\ }\bibfield  {title} {\bibinfo {title} {Mobility transition in a dynamic environment},\ }\href {https://doi.org/10.1088/1751-8113/47/25/255003} {\bibfield  {journal} {\bibinfo  {journal} {Journal of Physics A: Mathematical and Theoretical}\ }\textbf {\bibinfo {volume} {47}},\ \bibinfo {pages} {255003} (\bibinfo {year} {2014})}\BibitemShut {NoStop}%
\bibitem [{\citenamefont {Baiesi}\ \emph {et~al.}(2015)\citenamefont {Baiesi}, \citenamefont {Stella},\ and\ \citenamefont {Vanderzande}}]{baiesi2015role}%
  \BibitemOpen
  \bibfield  {author} {\bibinfo {author} {\bibfnamefont {M.}~\bibnamefont {Baiesi}}, \bibinfo {author} {\bibfnamefont {A.~L.}\ \bibnamefont {Stella}},\ and\ \bibinfo {author} {\bibfnamefont {C.}~\bibnamefont {Vanderzande}},\ }\bibfield  {title} {\bibinfo {title} {Role of trapping and crowding as sources of negative differential mobility},\ }\href {https://doi.org/10.1103/PhysRevE.92.042121} {\bibfield  {journal} {\bibinfo  {journal} {Physical Review E}\ }\textbf {\bibinfo {volume} {92}},\ \bibinfo {pages} {042121} (\bibinfo {year} {2015})}\BibitemShut {NoStop}%
\bibitem [{\citenamefont {Illien}\ \emph {et~al.}(2015)\citenamefont {Illien}, \citenamefont {B{\'e}nichou}, \citenamefont {Oshanin},\ and\ \citenamefont {Voituriez}}]{illien2015distribution}%
  \BibitemOpen
  \bibfield  {author} {\bibinfo {author} {\bibfnamefont {P.}~\bibnamefont {Illien}}, \bibinfo {author} {\bibfnamefont {O.}~\bibnamefont {B{\'e}nichou}}, \bibinfo {author} {\bibfnamefont {G.}~\bibnamefont {Oshanin}},\ and\ \bibinfo {author} {\bibfnamefont {R.}~\bibnamefont {Voituriez}},\ }\bibfield  {title} {\bibinfo {title} {Distribution of the position of a driven tracer in a hardcore lattice gas},\ }\href {https://doi.org/10.1088/1742-5468/2015/11/P11016} {\bibfield  {journal} {\bibinfo  {journal} {Journal of Statistical Mechanics: Theory and Experiment}\ }\textbf {\bibinfo {volume} {2015}},\ \bibinfo {pages} {P11016} (\bibinfo {year} {2015})}\BibitemShut {NoStop}%
\bibitem [{\citenamefont {Barkai}(2001)}]{barkai2001fractional}%
  \BibitemOpen
  \bibfield  {author} {\bibinfo {author} {\bibfnamefont {E.}~\bibnamefont {Barkai}},\ }\bibfield  {title} {\bibinfo {title} {Fractional fokker-planck equation, solution, and application},\ }\href {https://doi.org/10.1103/PhysRevE.63.046118} {\bibfield  {journal} {\bibinfo  {journal} {Physical Review E}\ }\textbf {\bibinfo {volume} {63}},\ \bibinfo {pages} {046118} (\bibinfo {year} {2001})}\BibitemShut {NoStop}%
\bibitem [{\citenamefont {Meerschaert}\ and\ \citenamefont {Scheffler}(2004)}]{meerschaert2004limit}%
  \BibitemOpen
  \bibfield  {author} {\bibinfo {author} {\bibfnamefont {M.~M.}\ \bibnamefont {Meerschaert}}\ and\ \bibinfo {author} {\bibfnamefont {H.-P.}\ \bibnamefont {Scheffler}},\ }\bibfield  {title} {\bibinfo {title} {Limit theorems for continuous-time random walks with infinite mean waiting times},\ }\href {https://doi.org/10.1239/jap/1091543414} {\bibfield  {journal} {\bibinfo  {journal} {Journal of Applied Probability}\ }\textbf {\bibinfo {volume} {41}},\ \bibinfo {pages} {623} (\bibinfo {year} {2004})}\BibitemShut {NoStop}%
\bibitem [{\citenamefont {Sokolov}\ and\ \citenamefont {Klafter}(2005)}]{sokolov2005diffusion}%
  \BibitemOpen
  \bibfield  {author} {\bibinfo {author} {\bibfnamefont {I.~M.}\ \bibnamefont {Sokolov}}\ and\ \bibinfo {author} {\bibfnamefont {J.}~\bibnamefont {Klafter}},\ }\bibfield  {title} {\bibinfo {title} {From diffusion to anomalous diffusion: a century after einstein’s brownian motion},\ }\href {https://doi.org/10.1063/1.1860472} {\bibfield  {journal} {\bibinfo  {journal} {Chaos: An Interdisciplinary Journal of Nonlinear Science}\ }\textbf {\bibinfo {volume} {15}} (\bibinfo {year} {2005})}\BibitemShut {NoStop}%
\bibitem [{\citenamefont {Yuste}\ and\ \citenamefont {Lindenberg}(2005)}]{yuste2005trapping}%
  \BibitemOpen
  \bibfield  {author} {\bibinfo {author} {\bibfnamefont {S.}~\bibnamefont {Yuste}}\ and\ \bibinfo {author} {\bibfnamefont {K.}~\bibnamefont {Lindenberg}},\ }\bibfield  {title} {\bibinfo {title} {Trapping reactions with subdiffusive traps and particles characterized by different anomalous diffusion exponents},\ }\href {https://doi.org/10.1103/PhysRevE.72.061103} {\bibfield  {journal} {\bibinfo  {journal} {Physical Review E}\ }\textbf {\bibinfo {volume} {72}},\ \bibinfo {pages} {061103} (\bibinfo {year} {2005})}\BibitemShut {NoStop}%
\bibitem [{\citenamefont {Saichev}\ and\ \citenamefont {Zaslavsky}(1997)}]{saichev1997fractional}%
  \BibitemOpen
  \bibfield  {author} {\bibinfo {author} {\bibfnamefont {A.~I.}\ \bibnamefont {Saichev}}\ and\ \bibinfo {author} {\bibfnamefont {G.~M.}\ \bibnamefont {Zaslavsky}},\ }\bibfield  {title} {\bibinfo {title} {Fractional kinetic equations: solutions and applications},\ }\href {https://doi.org/10.1063/1.166272} {\bibfield  {journal} {\bibinfo  {journal} {Chaos: An Interdisciplinary Journal of Nonlinear Science}\ }\textbf {\bibinfo {volume} {7}},\ \bibinfo {pages} {753} (\bibinfo {year} {1997})}\BibitemShut {NoStop}%
\bibitem [{\citenamefont {Ballentine}(2014)}]{ballentine2014quantum}%
  \BibitemOpen
  \bibfield  {author} {\bibinfo {author} {\bibfnamefont {L.~E.}\ \bibnamefont {Ballentine}},\ }\href {https://books.google.com/books?hl=en&lr=&id=Yi48DQAAQBAJ&oi=fnd&pg=PR5&dq=Quantum+mechanics:+a+modern+development&ots=UFE_D6hQ0L&sig=IjW1jiBdjRX8-UnrZBjZfAqzfQo} {\emph {\bibinfo {title} {Quantum mechanics: a modern development}}}\ (\bibinfo  {publisher} {World Scientific Publishing Company},\ \bibinfo {year} {2014})\BibitemShut {NoStop}%
\bibitem [{\citenamefont {Illien}\ \emph {et~al.}(2018)\citenamefont {Illien}, \citenamefont {B{\'e}nichou}, \citenamefont {Oshanin}, \citenamefont {Sarracino},\ and\ \citenamefont {Voituriez}}]{illien2018nonequilibrium}%
  \BibitemOpen
  \bibfield  {author} {\bibinfo {author} {\bibfnamefont {P.}~\bibnamefont {Illien}}, \bibinfo {author} {\bibfnamefont {O.}~\bibnamefont {B{\'e}nichou}}, \bibinfo {author} {\bibfnamefont {G.}~\bibnamefont {Oshanin}}, \bibinfo {author} {\bibfnamefont {A.}~\bibnamefont {Sarracino}},\ and\ \bibinfo {author} {\bibfnamefont {R.}~\bibnamefont {Voituriez}},\ }\bibfield  {title} {\bibinfo {title} {Nonequilibrium fluctuations and enhanced diffusion of a driven particle in a dense environment},\ }\href {https://doi.org/10.1103/PhysRevLett.120.200606} {\bibfield  {journal} {\bibinfo  {journal} {Physical Review Letters}\ }\textbf {\bibinfo {volume} {120}},\ \bibinfo {pages} {200606} (\bibinfo {year} {2018})}\BibitemShut {NoStop}%
\bibitem [{\citenamefont {Akimoto}\ \emph {et~al.}(2018)\citenamefont {Akimoto}, \citenamefont {Barkai},\ and\ \citenamefont {Saito}}]{akimoto2018non}%
  \BibitemOpen
  \bibfield  {author} {\bibinfo {author} {\bibfnamefont {T.}~\bibnamefont {Akimoto}}, \bibinfo {author} {\bibfnamefont {E.}~\bibnamefont {Barkai}},\ and\ \bibinfo {author} {\bibfnamefont {K.}~\bibnamefont {Saito}},\ }\bibfield  {title} {\bibinfo {title} {Non-self-averaging behaviors and ergodicity in quenched trap models with finite system sizes},\ }\href {https://doi.org/10.1103/PhysRevE.97.052143} {\bibfield  {journal} {\bibinfo  {journal} {Physical Review E}\ }\textbf {\bibinfo {volume} {97}},\ \bibinfo {pages} {052143} (\bibinfo {year} {2018})}\BibitemShut {NoStop}%
\bibitem [{\citenamefont {Akimoto}\ and\ \citenamefont {Saito}(2020)}]{akimoto2020trace}%
  \BibitemOpen
  \bibfield  {author} {\bibinfo {author} {\bibfnamefont {T.}~\bibnamefont {Akimoto}}\ and\ \bibinfo {author} {\bibfnamefont {K.}~\bibnamefont {Saito}},\ }\bibfield  {title} {\bibinfo {title} {Trace of anomalous diffusion in a biased quenched trap model},\ }\href {https://doi.org/10.1103/PhysRevE.101.042133} {\bibfield  {journal} {\bibinfo  {journal} {Physical Review E}\ }\textbf {\bibinfo {volume} {101}},\ \bibinfo {pages} {042133} (\bibinfo {year} {2020})}\BibitemShut {NoStop}%
\bibitem [{\citenamefont {Burov}\ and\ \citenamefont {Barkai}(2011)}]{burov2011time}%
  \BibitemOpen
  \bibfield  {author} {\bibinfo {author} {\bibfnamefont {S.}~\bibnamefont {Burov}}\ and\ \bibinfo {author} {\bibfnamefont {E.}~\bibnamefont {Barkai}},\ }\bibfield  {title} {\bibinfo {title} {Time transformation for random walks in the quenched trap model},\ }\href {https://doi.org/10.1103/PhysRevLett.106.140602} {\bibfield  {journal} {\bibinfo  {journal} {Physical Review Letters}\ }\textbf {\bibinfo {volume} {106}},\ \bibinfo {pages} {140602} (\bibinfo {year} {2011})}\BibitemShut {NoStop}%
\bibitem [{\citenamefont {Burov}(2017)}]{burov2017quenched}%
  \BibitemOpen
  \bibfield  {author} {\bibinfo {author} {\bibfnamefont {S.}~\bibnamefont {Burov}},\ }\bibfield  {title} {\bibinfo {title} {From quenched disorder to continuous time random walk},\ }\href {https://doi.org/10.1103/PhysRevE.96.050103} {\bibfield  {journal} {\bibinfo  {journal} {Physical Review E}\ }\textbf {\bibinfo {volume} {96}},\ \bibinfo {pages} {050103} (\bibinfo {year} {2017})}\BibitemShut {NoStop}%
\bibitem [{\citenamefont {Shafir}\ and\ \citenamefont {Burov}(2024)}]{shafir2024disorder}%
  \BibitemOpen
  \bibfield  {author} {\bibinfo {author} {\bibfnamefont {D.}~\bibnamefont {Shafir}}\ and\ \bibinfo {author} {\bibfnamefont {S.}~\bibnamefont {Burov}},\ }\bibfield  {title} {\bibinfo {title} {Disorder-induced anomalous mobility enhancement in confined geometries},\ }\href {https://doi.org/10.1103/PhysRevLett.133.037101} {\bibfield  {journal} {\bibinfo  {journal} {Physical Review Letters}\ }\textbf {\bibinfo {volume} {133}},\ \bibinfo {pages} {037101} (\bibinfo {year} {2024})}\BibitemShut {NoStop}%
\bibitem [{\citenamefont {Akimoto}\ \emph {et~al.}(2016)\citenamefont {Akimoto}, \citenamefont {Barkai},\ and\ \citenamefont {Saito}}]{akimoto2016universal}%
  \BibitemOpen
  \bibfield  {author} {\bibinfo {author} {\bibfnamefont {T.}~\bibnamefont {Akimoto}}, \bibinfo {author} {\bibfnamefont {E.}~\bibnamefont {Barkai}},\ and\ \bibinfo {author} {\bibfnamefont {K.}~\bibnamefont {Saito}},\ }\bibfield  {title} {\bibinfo {title} {Universal fluctuations of single-particle diffusivity in a quenched environment},\ }\href {https://doi.org/10.1103/PhysRevLett.117.180602} {\bibfield  {journal} {\bibinfo  {journal} {Physical review letters}\ }\textbf {\bibinfo {volume} {117}},\ \bibinfo {pages} {180602} (\bibinfo {year} {2016})}\BibitemShut {NoStop}%
\bibitem [{\citenamefont {Weiss}\ and\ \citenamefont {Weiss}(1994)}]{weiss1994aspects}%
  \BibitemOpen
  \bibfield  {author} {\bibinfo {author} {\bibfnamefont {G.~H.}\ \bibnamefont {Weiss}}\ and\ \bibinfo {author} {\bibfnamefont {G.~H.}\ \bibnamefont {Weiss}},\ }\href {https://cir.nii.ac.jp/crid/1130000796986494208} {\emph {\bibinfo {title} {Aspects and applications of the random walk}}}\ (\bibinfo  {publisher} {Elsevier Science \& Technology},\ \bibinfo {year} {1994})\BibitemShut {NoStop}%
\bibitem [{\citenamefont {Hughes}(1995)}]{hughes1995random}%
  \BibitemOpen
  \bibfield  {author} {\bibinfo {author} {\bibfnamefont {B.~D.}\ \bibnamefont {Hughes}},\ }\href {https://doi.org/10.1093/oso/9780198537892.001.000} {\emph {\bibinfo {title} {Random walks and random environments: random walks}}},\ Vol.~\bibinfo {volume} {1}\ (\bibinfo  {publisher} {Oxford University Press},\ \bibinfo {year} {1995})\BibitemShut {NoStop}%
\bibitem [{\citenamefont {Klafter}\ and\ \citenamefont {Sokolov}(2011)}]{klafter2011first}%
  \BibitemOpen
  \bibfield  {author} {\bibinfo {author} {\bibfnamefont {J.}~\bibnamefont {Klafter}}\ and\ \bibinfo {author} {\bibfnamefont {I.~M.}\ \bibnamefont {Sokolov}},\ }\href {https://doi.org/10.1093/acprof:oso/9780199234868.001.0001} {\emph {\bibinfo {title} {First steps in random walks: from tools to applications}}}\ (\bibinfo  {publisher} {OUP Oxford},\ \bibinfo {year} {2011})\BibitemShut {NoStop}%
\bibitem [{\citenamefont {Brummelhuis}\ and\ \citenamefont {Hilhorst}(1988)}]{brummelhuis1988single}%
  \BibitemOpen
  \bibfield  {author} {\bibinfo {author} {\bibfnamefont {M.}~\bibnamefont {Brummelhuis}}\ and\ \bibinfo {author} {\bibfnamefont {H.}~\bibnamefont {Hilhorst}},\ }\bibfield  {title} {\bibinfo {title} {Single-vacancy induced motion of a tracer particle in a two-dimensional lattice gas},\ }\href {https://doi.org/10.1007/BF01011556} {\bibfield  {journal} {\bibinfo  {journal} {Journal of Statistical Physics}\ }\textbf {\bibinfo {volume} {53}},\ \bibinfo {pages} {249} (\bibinfo {year} {1988})}\BibitemShut {NoStop}%
\bibitem [{\citenamefont {Nieuwenhuizen}\ \emph {et~al.}(1986)\citenamefont {Nieuwenhuizen}, \citenamefont {Van~Velthoven},\ and\ \citenamefont {Ernst}}]{nieuwenhuizen1986diffusion}%
  \BibitemOpen
  \bibfield  {author} {\bibinfo {author} {\bibfnamefont {T.~M.}\ \bibnamefont {Nieuwenhuizen}}, \bibinfo {author} {\bibfnamefont {P.}~\bibnamefont {Van~Velthoven}},\ and\ \bibinfo {author} {\bibfnamefont {M.}~\bibnamefont {Ernst}},\ }\bibfield  {title} {\bibinfo {title} {Diffusion and long-time tails in a two-dimensional site-percolation model},\ }\href {https://doi.org/10.1103/PhysRevLett.57.2477} {\bibfield  {journal} {\bibinfo  {journal} {Physical Review Letters}\ }\textbf {\bibinfo {volume} {57}},\ \bibinfo {pages} {2477} (\bibinfo {year} {1986})}\BibitemShut {NoStop}%
\bibitem [{\citenamefont {Ernst}\ and\ \citenamefont {Weyland}(1971)}]{ernst1971long}%
  \BibitemOpen
  \bibfield  {author} {\bibinfo {author} {\bibfnamefont {M.}~\bibnamefont {Ernst}}\ and\ \bibinfo {author} {\bibfnamefont {A.}~\bibnamefont {Weyland}},\ }\bibfield  {title} {\bibinfo {title} {Long time behaviour of the velocity auto-correlation function in a lorentz gas},\ }\href {https://doi.org/10.1016/0375-9601(71)90987-X} {\bibfield  {journal} {\bibinfo  {journal} {Physics Letters A}\ }\textbf {\bibinfo {volume} {34}},\ \bibinfo {pages} {39} (\bibinfo {year} {1971})}\BibitemShut {NoStop}%
\bibitem [{\citenamefont {Nieuwenhuizen}\ \emph {et~al.}(1987)\citenamefont {Nieuwenhuizen}, \citenamefont {Van~Velthoven},\ and\ \citenamefont {Ernst}}]{nieuwenhuizen1987density}%
  \BibitemOpen
  \bibfield  {author} {\bibinfo {author} {\bibfnamefont {T.~M.}\ \bibnamefont {Nieuwenhuizen}}, \bibinfo {author} {\bibfnamefont {P.}~\bibnamefont {Van~Velthoven}},\ and\ \bibinfo {author} {\bibfnamefont {M.}~\bibnamefont {Ernst}},\ }\bibfield  {title} {\bibinfo {title} {Density expansion of transport properties on 2d site-disordered lattices: I. general theory},\ }\href {https://doi.org/10.1088/0305-4470/20/12/044} {\bibfield  {journal} {\bibinfo  {journal} {Journal of Physics A: Mathematical and General}\ }\textbf {\bibinfo {volume} {20}},\ \bibinfo {pages} {4001} (\bibinfo {year} {1987})}\BibitemShut {NoStop}%
\bibitem [{\citenamefont {Montroll}\ and\ \citenamefont {Scher}(1973)}]{montroll1973random}%
  \BibitemOpen
  \bibfield  {author} {\bibinfo {author} {\bibfnamefont {E.~W.}\ \bibnamefont {Montroll}}\ and\ \bibinfo {author} {\bibfnamefont {H.}~\bibnamefont {Scher}},\ }\bibfield  {title} {\bibinfo {title} {Random walks on lattices. iv. continuous-time walks and influence of absorbing boundaries},\ }\href {https://doi.org/10.1007/BF01016843} {\bibfield  {journal} {\bibinfo  {journal} {Journal of Statistical Physics}\ }\textbf {\bibinfo {volume} {9}},\ \bibinfo {pages} {101} (\bibinfo {year} {1973})}\BibitemShut {NoStop}%
\bibitem [{\citenamefont {Frenkel}(1987)}]{frenkel1987velocity}%
  \BibitemOpen
  \bibfield  {author} {\bibinfo {author} {\bibfnamefont {D.}~\bibnamefont {Frenkel}},\ }\bibfield  {title} {\bibinfo {title} {Velocity auto-correlation functions in a 2d lattice lorentz gas: Comparison of theory and computer simulation},\ }\href {https://doi.org/10.1016/0375-9601(87)90482-8} {\bibfield  {journal} {\bibinfo  {journal} {Physics Letters A}\ }\textbf {\bibinfo {volume} {121}},\ \bibinfo {pages} {385} (\bibinfo {year} {1987})}\BibitemShut {NoStop}%
\bibitem [{\citenamefont {Rusch}\ \emph {et~al.}(2024)\citenamefont {Rusch}, \citenamefont {Franosch},\ and\ \citenamefont {Jung}}]{rusch2024noise}%
  \BibitemOpen
  \bibfield  {author} {\bibinfo {author} {\bibfnamefont {R.}~\bibnamefont {Rusch}}, \bibinfo {author} {\bibfnamefont {T.}~\bibnamefont {Franosch}},\ and\ \bibinfo {author} {\bibfnamefont {G.}~\bibnamefont {Jung}},\ }\bibfield  {title} {\bibinfo {title} {Noise-cancellation algorithm for simulations of brownian particles},\ }\href {https://doi.org/10.1103/PhysRevE.109.015303} {\bibfield  {journal} {\bibinfo  {journal} {Physical Review E}\ }\textbf {\bibinfo {volume} {109}},\ \bibinfo {pages} {015303} (\bibinfo {year} {2024})}\BibitemShut {NoStop}%
\bibitem [{\citenamefont {Chambers}\ \emph {et~al.}(1976)\citenamefont {Chambers}, \citenamefont {Mallows},\ and\ \citenamefont {Stuck}}]{chambers1976method}%
  \BibitemOpen
  \bibfield  {author} {\bibinfo {author} {\bibfnamefont {J.~M.}\ \bibnamefont {Chambers}}, \bibinfo {author} {\bibfnamefont {C.~L.}\ \bibnamefont {Mallows}},\ and\ \bibinfo {author} {\bibfnamefont {B.}~\bibnamefont {Stuck}},\ }\bibfield  {title} {\bibinfo {title} {A method for simulating stable random variables},\ }\href {https://doi.org/10.1080/01621459.1976.10480344} {\bibfield  {journal} {\bibinfo  {journal} {Journal of the American Statistical Association}\ }\textbf {\bibinfo {volume} {71}},\ \bibinfo {pages} {340} (\bibinfo {year} {1976})}\BibitemShut {NoStop}%
\end{thebibliography}%

\end{document}